\documentclass{elsart}

\usepackage[square,comma]{natbib}
\usepackage{graphicx}
\usepackage{pxfonts}
\usepackage{lineno}

\usepackage{amssymb}
\usepackage{footnote}
\usepackage{multirow}
\usepackage{varwidth}
\usepackage{array}

\journal{}

\begin{document}

\thispagestyle{empty}
\begin{Large}
\textbf{DEUTSCHES ELEKTRONEN-SYNCHROTRON}

\textbf{\large{Ein Forschungszentrum der Helmholtz-Gemeinschaft}\\}
\end{Large}

DESY 13-135

July 2013

\begin{eqnarray}
\nonumber
\end{eqnarray}
\begin{center}
\begin{Large}
\textbf{Purified SASE undulator configuration to enhance the
performance of the soft x-ray beamline at the European XFEL}
\end{Large}
\begin{eqnarray}
\nonumber &&\cr \nonumber
\end{eqnarray}

\begin{large}
Svitozar Serkez$^a$, Vitali Kocharyan$^a$, Evgeni Saldin$^a$, Igor
Zagorodnov$^a$, Gianluca Geloni$^b$
\end{large}

\textsl{\\$^a$Deutsches Elektronen-Synchrotron DESY, Hamburg}
\begin{large}

\end{large}
\textsl{\\$^b$European XFEL GmbH, Hamburg}
\begin{large}

\end{large}

\begin{eqnarray}
\nonumber
\end{eqnarray}
\begin{eqnarray}
\nonumber
\end{eqnarray}
ISSN 0418-9833
\begin{eqnarray}
\nonumber
\end{eqnarray}
\begin{large}
\textbf{NOTKESTRASSE 85 - 22607 HAMBURG}
\end{large}
\end{center}
\clearpage
\newpage

\begin{frontmatter}



\title{Purified SASE undulator configuration to enhance the performance of the soft x-ray beamline at the European XFEL}


\author[DESY]{Svitozar Serkez \thanksref{corr},}
\thanks[corr]{Corresponding Author. E-mail address: svitozar.serkez@desy.de}
\author[DESY]{Vitali Kocharyan,}
\author[DESY]{Evgeni Saldin,}
\author[DESY]{Igor Zagorodnov,}
\author[XFEL]{Gianluca Geloni,}

\address[DESY]{Deutsches Elektronen-Synchrotron (DESY), Hamburg, Germany}
\address[XFEL]{European XFEL GmbH, Hamburg, Germany}

\begin{abstract}
The purified SASE (pSASE) undulator configuration recently proposed
at SLAC promises an increase in the output spectral density of
XFELs. In this article we study a straightforward implementation of
this configuration for the soft x-ray beamline at the European XFEL.
A few undulator cells, resonant at a subharmonic of the FEL
radiation, are used in the middle of the exponential regime to
amplify the radiation, while simultaneously reducing the FEL
bandwidth. Based on start-to-end simulations, we show that with the
proposed configuration the spectral density in the photon energy
range between 1.3 keV and 3 keV can be enhanced of an order of
magnitude compared to the baseline mode of operation. This option
can be implemented into the tunable-gap SASE3 baseline undulator
without additional hardware, and it is complementary to the
self-seeding option with grating monochromator proposed for the same
undulator line, which can cover the photon energy range between
about 0.26 keV and 1 keV.
\end{abstract}

%
%
%
\end{frontmatter}



\section{\label{sec:intro} Introduction}

This article discusses the potential for enhancing the capability of
the soft x-ray (SASE3) beamline at the European XFEL which will be
operated in the photon energy range between 0.26 keV and at least 3
keV. A high level of longitudinal coherence is the key to upgrade
the baseline performance. Self-seeding is a promising approach to
significantly narrow the SASE bandwidth and to produce nearly
transform-limited pulses \cite{SELF}-\cite{ASYM}. The implementation
of this method in the soft x-ray wavelength range necessarily
involves gratings as dispersive elements. A grating monochromator,
which can be installed in the SASE3 undulator without perturbing the
electron focusing system can cover the spectral range between about
0.26 keV and 1 keV \cite{FENG3}-\cite{GRAT}.

One of the main technical problems for self-seeding designers is to
provide a high level of longitudinal coherence in the photon energy
range between 1 keV and 3 keV. In this range, proposals exist to
narrow the SASE bandwidth at the European XFEL  by combining
self-seeding and fresh bunch techniques. However, this requires
installing additional hardware in the undulator system
\cite{BIO1,BIO3}. Here we explore a simpler method to reach
practically the same result without further changes in the undulator
system. The solution is based in essence on the purified SASE
(pSASE) technique recently proposed at SLAC \cite{PSASE}, and
naturally exploits the gap tunability of the SASE3 undulator.

In the pSASE configuration, a few undulator cells resonant at a
subharmonic of the FEL radiation, called altogether the
"slippage-boosted section", are used in the high-gain linear regime
to reduce the SASE bandwidth. The final characteristics of a pSASE
source are a compromise between high output power, which can be
reached with a conventional SASE undulator source resonant at the
target wavelength, and narrow bandwidth, which can be reached with
harmonic lasing \cite{BONI}-\cite{SCHN}.

Here we demonstrate that it is possible to cover the energy range
between 1.3 keV and 3 keV using the nominal European XFEL electron
beam parameters, and to reduce the SASE bandwidth by a factor 5,
still having the same output power as in the baseline SASE regime.
Note that the slippage-boosted section is tuned to a subharmonic
(the fifth, or the seventh) of the FEL radiation. Therefore, the
choice of the lowest pSASE photon energy considered in this article,
1.3 keV, is dictated by the minimal photon energy (0.26 keV) that
can be reached in the conventional SASE regime.

\section{\label{sec:desc} Setup description}

\begin{figure}[tb]
\begin{center}
\includegraphics[clip,width=1.0\textwidth]{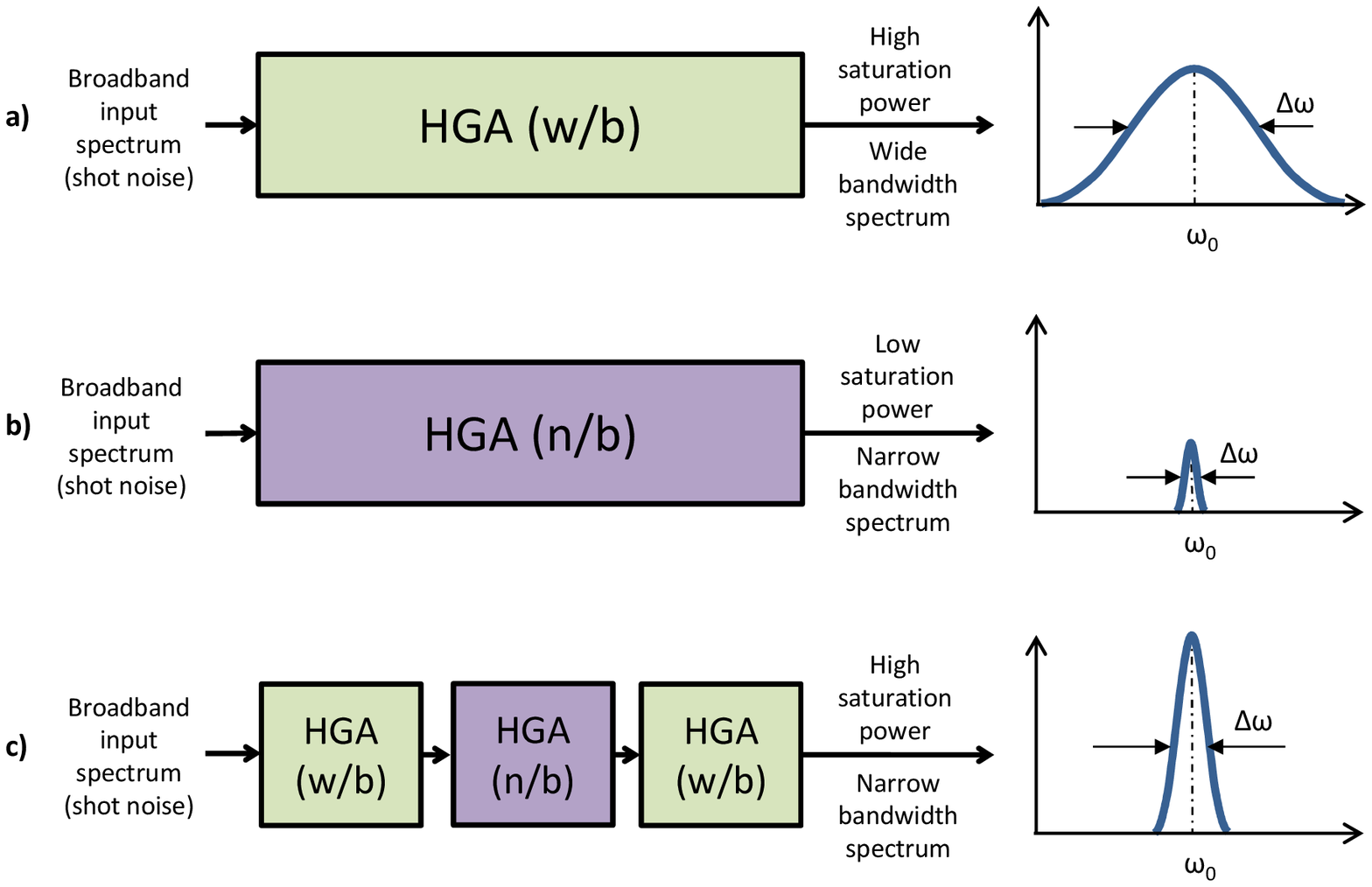}
\end{center}
\caption{Interpretation of the pSASE concept as a compromise between
the high output power of a conventional SASE source and a
narrow-bandwidth harmonic-lasing SASE source. a) Model of a
conventional SASE source operating at the fundamental harmonic of
the undulator. b) Model of a SASE source operating at a higher
harmonic (5th-7th) of the fundamental. c) Model of a pSASE source. }
\label{compro}
\end{figure}
\begin{figure}[tb]
\begin{center}
\includegraphics[clip,width=1.0\textwidth]{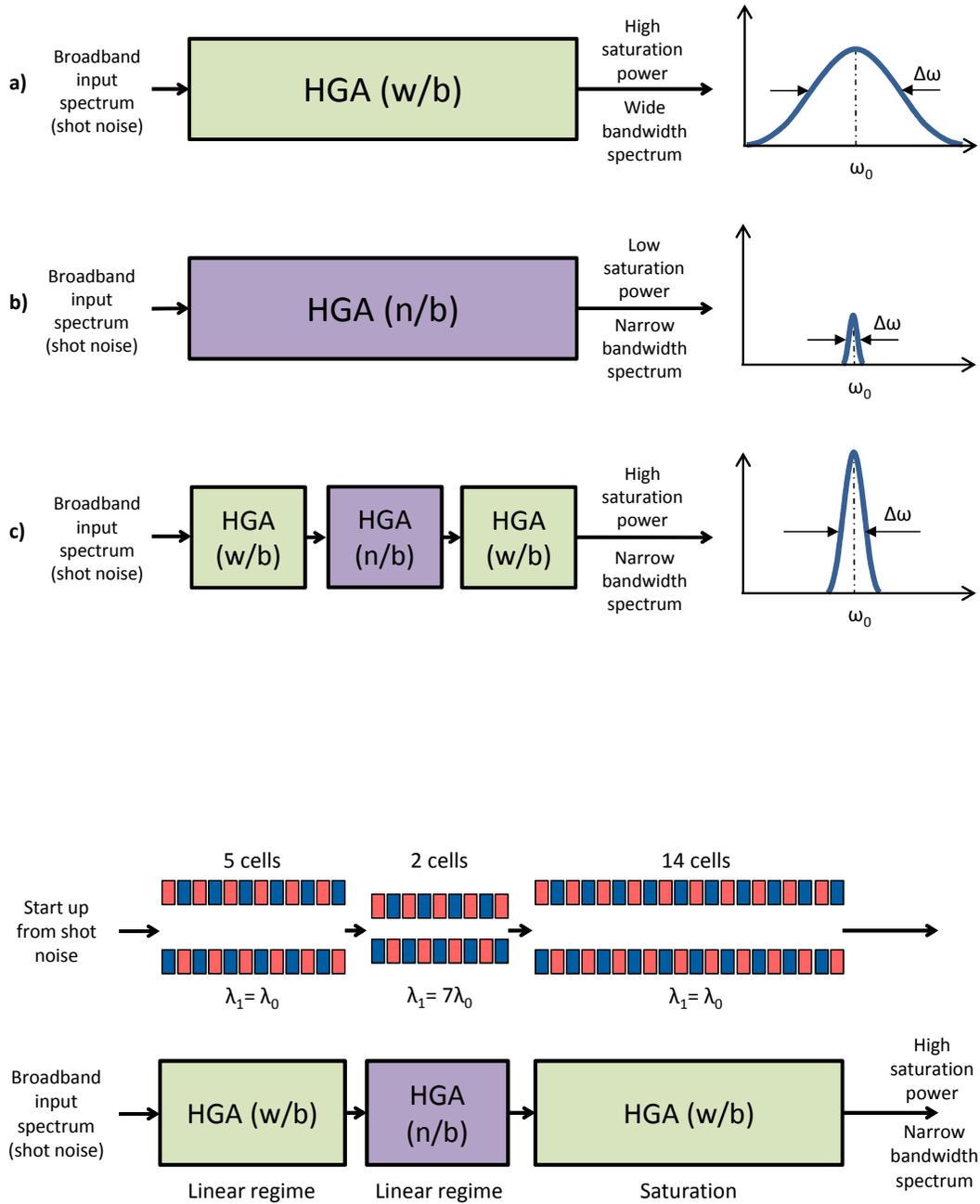}
\end{center}
\caption{The actual pSASE undulator configuration proposed for the
SASE3 beamline, which is expected to operate in the photon energy
range between $1.3$ keV and $3$ keV. } \label{pSASE}
\end{figure}
Let us consider the SASE3 undulator configured in pSASE mode as
proposed in \cite{PSASE}\footnote{The first mention of what might be
called a pSASE undulator configuration appears, to the best of our
knowledge, in \cite{SCHN}. Half a year later, the idea was much more
fleshed out in \cite{PSASE}. Using the LCLS-II parameters as an
example, in \cite{PSASE} was shown that, with the proposed pSASE
configuration, the temporal coherence and spectral brightness of a
SASE FEL can be significantly enhanced.}. The operation of this
setup can be understood from Fig. \ref{compro}. Suppose that two
high-gain FEL amplifiers based on the same tunable-gap undulator
start from shot noise and generate SASE radiation at the same target
frequency. The undulator system of the first amplifier is tuned to
operate at this target frequency as the fundamental harmonic, Fig.
\ref{compro}(a). The undulator system of the second amplifier in
Fig. \ref{compro}(b) is tuned, instead, to operate at the same
target frequency lasing at an harmonic of the fundamental, a mode of
operation called "harmonic lasing" \cite{BONI}-\cite{SCHN}. Harmonic
lasing is possible once the growth of the radiation at the
fundamental frequency, which is now a sub-harmonic of the target
frequency, is suppressed, for example by properly tuning the phase
shifters after each undulator section. The first FEL amplifier,
operating at the fundamental harmonic, is capable of generating SASE
radiation with a high-power level at saturation, but with a
relatively wide spectrum. The second amplifier, operating in
harmonic-lasing mode produces a narrower bandwidth instead, but also
a lower level saturation power \cite{SCHN}. Let us now consider a
third configuration where, instead of using harmonic lasing up to
saturation, one exploits it as an active filter in the linear
regime. This configuration, illustrated in Fig. \ref{compro}(c) is a
pSASE configuration. In this case, at saturation one can reach both
a high-power level and a narrow spectrum. We will qualitatively
explain the reasons for these effects further on in this section.

A schematic layout of the proposed pSASE configuration for the SASE3
undulator at the European XFEL is illustrated in Fig. \ref{pSASE}.
Following \cite{PSASE}, the undulator consists of three parts, U1,
U2, and U3. The first undulator part U1 is resonant at the target
FEL frequency $\omega_1 = \omega_0$, and is used to produce standard
SASE radiation in pulses with central frequency $\omega_0$. The
length of U1 is chosen in such a way that U1 operates in the linear
high-gain regime. The SASE radiation and the electron beam then
enter the second undulator part U2, which is made resonant at the
nth subharmonic of the target FEL frequency $\omega_1 = \omega_0/n$
by properly increasing the undulator parameter. In the
slippage-boosted section U2 the SASE radiation is amplified through
the harmonic interaction, while its bandwidth is simultaneously
reduced.

The narrow bandwidth radiation is finally amplified to saturation in
the last undulator part, U3, which is resonant again at frequency
$\omega_1 = \omega_0$ as in U1. With this configuration, the FEL
reaches a high-power level with significantly reduced radiation
bandwidth. The main purpose of U2 is indeed to reduce the SASE
bandwidth. Operating U3 at the same frequency as U1 makes the
saturation power of the SASE FEL essentially the same as in the
nominal SASE configuration, which allows one to increase the output
spectral brightness.

We optimized our setup based on start-to-end simulations for the
nominal electron beam with 100 pC charge at $10.5$ GeV. At the
longitudinal position corresponding to the maximum current value,
$5$ kA, the electron bunch has a normalized emittance of about
$0.2~\mu$m, and an energy spread of about $2$ MeV \cite{ZAGO}.
Simulations were performed with the help of the Genesis code
\cite{GENE}. We first modeled the performance of the high-gain FEL
amplifiers described in Fig. \ref{compro}(a) and Fig.
\ref{compro}(b) in the steady state regime. This allowed us to
calculate the gain curve of the amplifiers in these two
configurations.

\begin{figure}[tb]
\includegraphics[clip,width=0.5\textwidth]{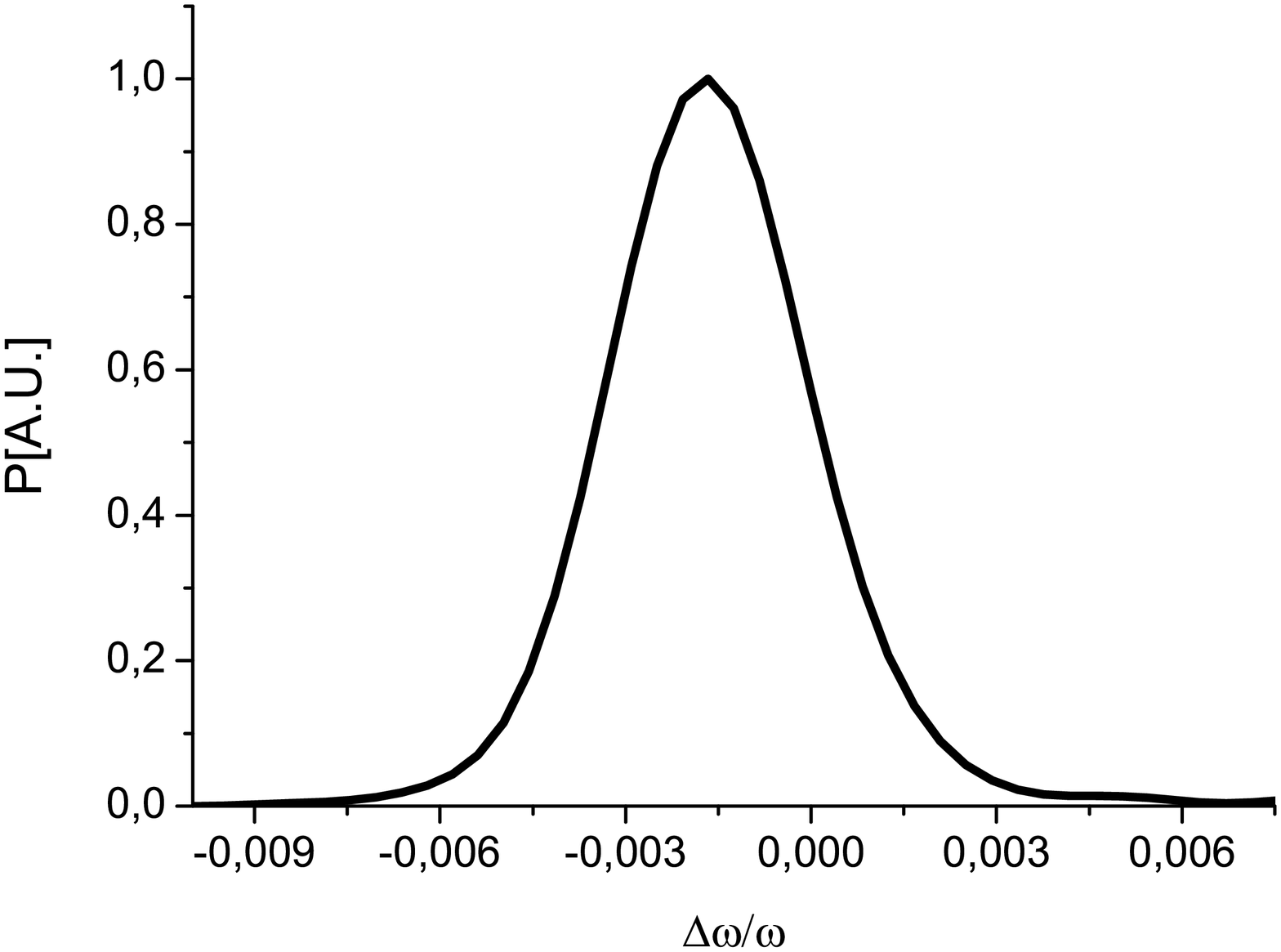}
\includegraphics[clip,width=0.5\textwidth]{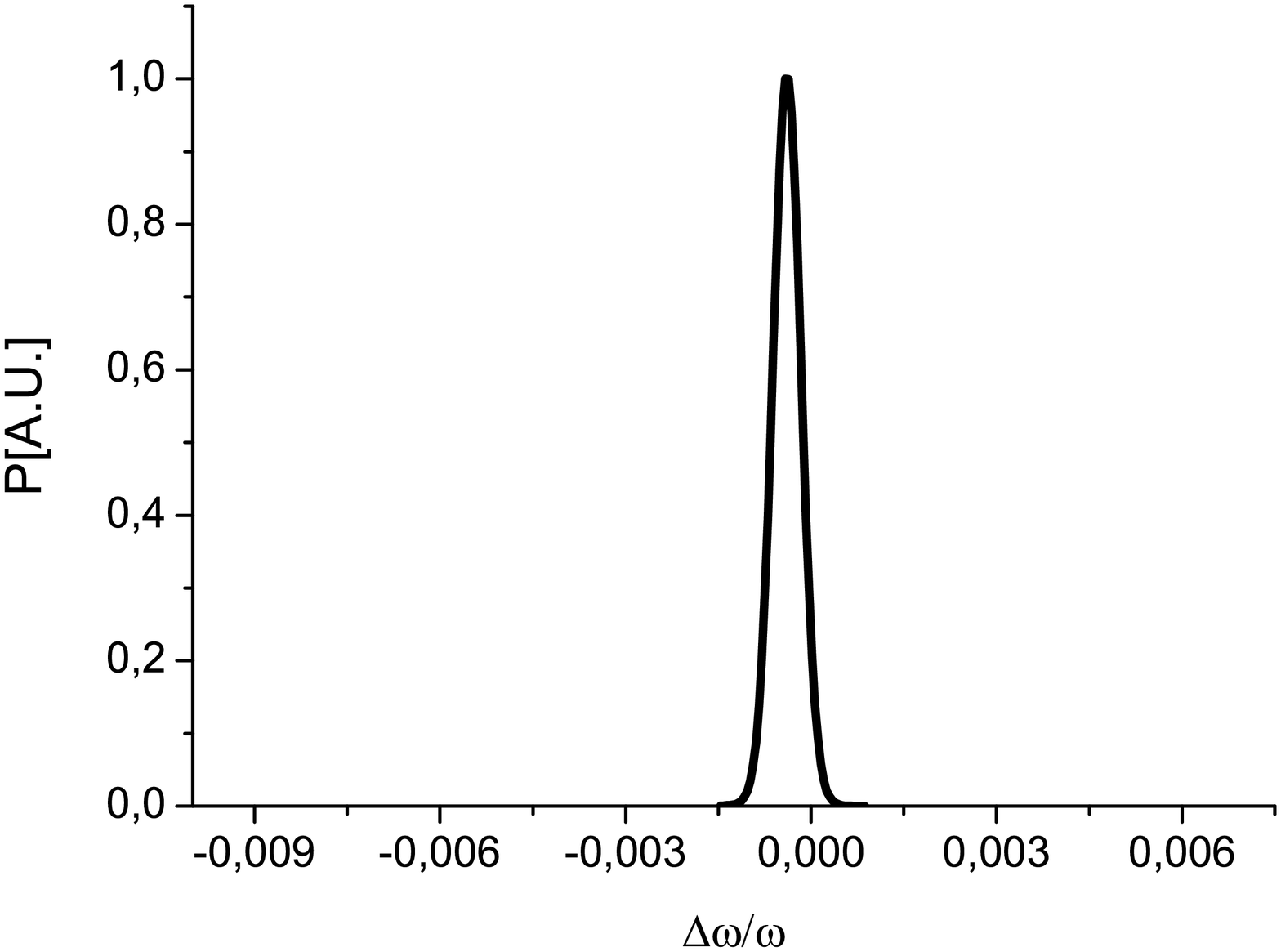}
\caption{The FEL configurations considered here refer to the SASE3
undulator line operating at $10.5$ GeV. Here the target photon
energy is 2 keV. Electron beam characteristics (normalized emittance
of about $0.2~\mu$m, energy spread of about $2$ MeV, peak current of
about $5$ kA) refer to the longitudinal position, inside the 0.1 nC
bunch, corresponding to the maximum current value (see section 3)
Left: Power gain versus the frequency. The curve is the result of
numerical simulations. The FEL amplifier operates at the fundamental
mode in the steady state, high-gain linear regime. Right: Power gain
versus frequency for the 7th harmonic lasing case. } \label{Gcomp}
\end{figure}
Fig. \ref{Gcomp} shows the comparison of gain curves for these two
modes of operation at a photon energy of 2 keV. For the harmonic
lasing case, the right plot in Fig. \ref{Gcomp}, the SASE3 undulator
is tuned to the 7th harmonic. The undulator $K$ value is tuned to
produce radiation with the same frequency as in the case of lasing
at the fundamental, as is illustrated in the left plot of the same
figure. These simulations clearly demonstrate the possibility of
producing narrow bandwidth radiation in the harmonic lasing mode.
When tuned to the 7th harmonic of the undulator, the FEL amplifier
is characterized by a bandwidth that is about $5$ times narrower
than in the case of lasing at the fundamental.

\begin{figure}[tb]
\includegraphics[clip,width=0.5\textwidth]{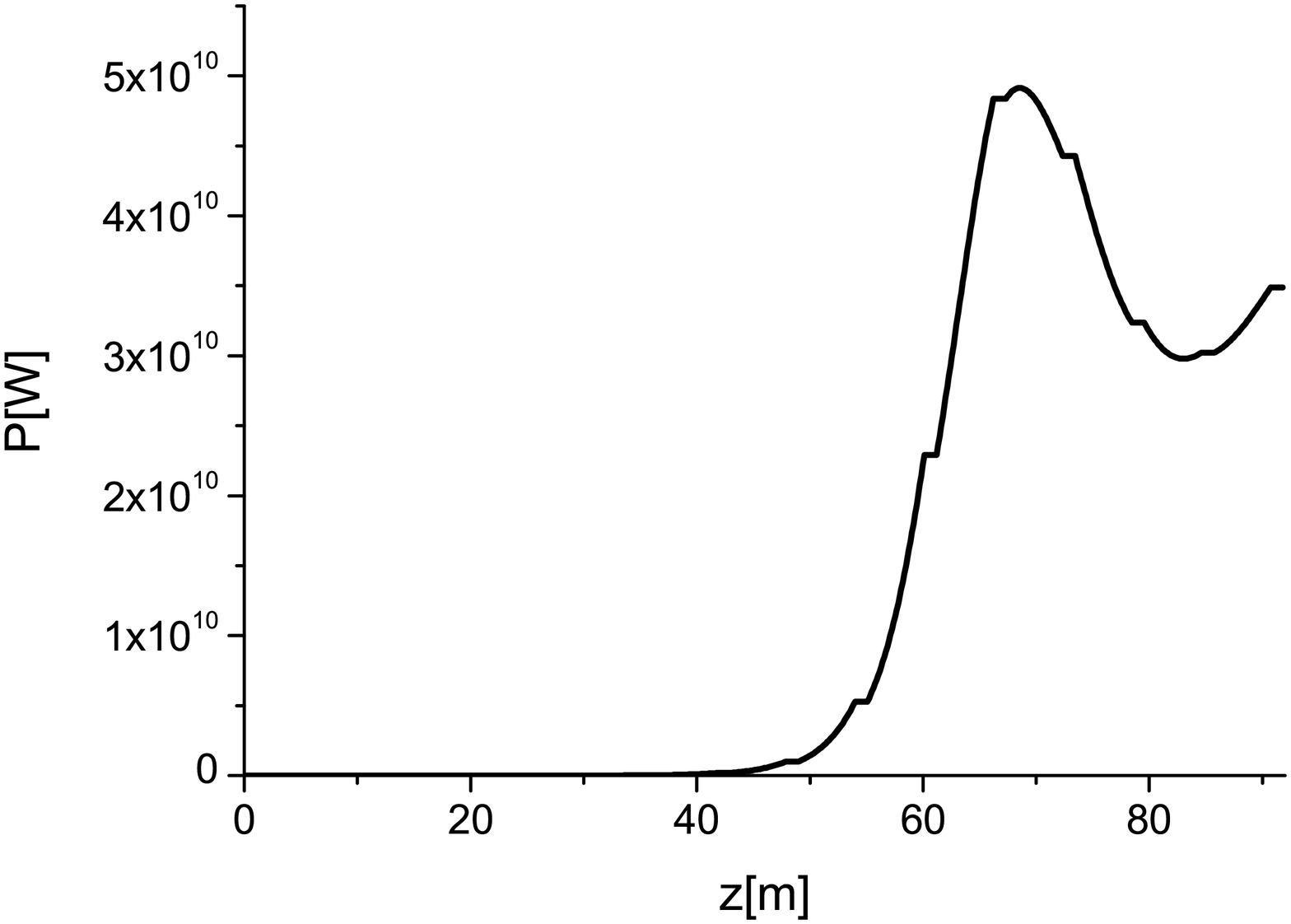}
\includegraphics[clip,width=0.5\textwidth]{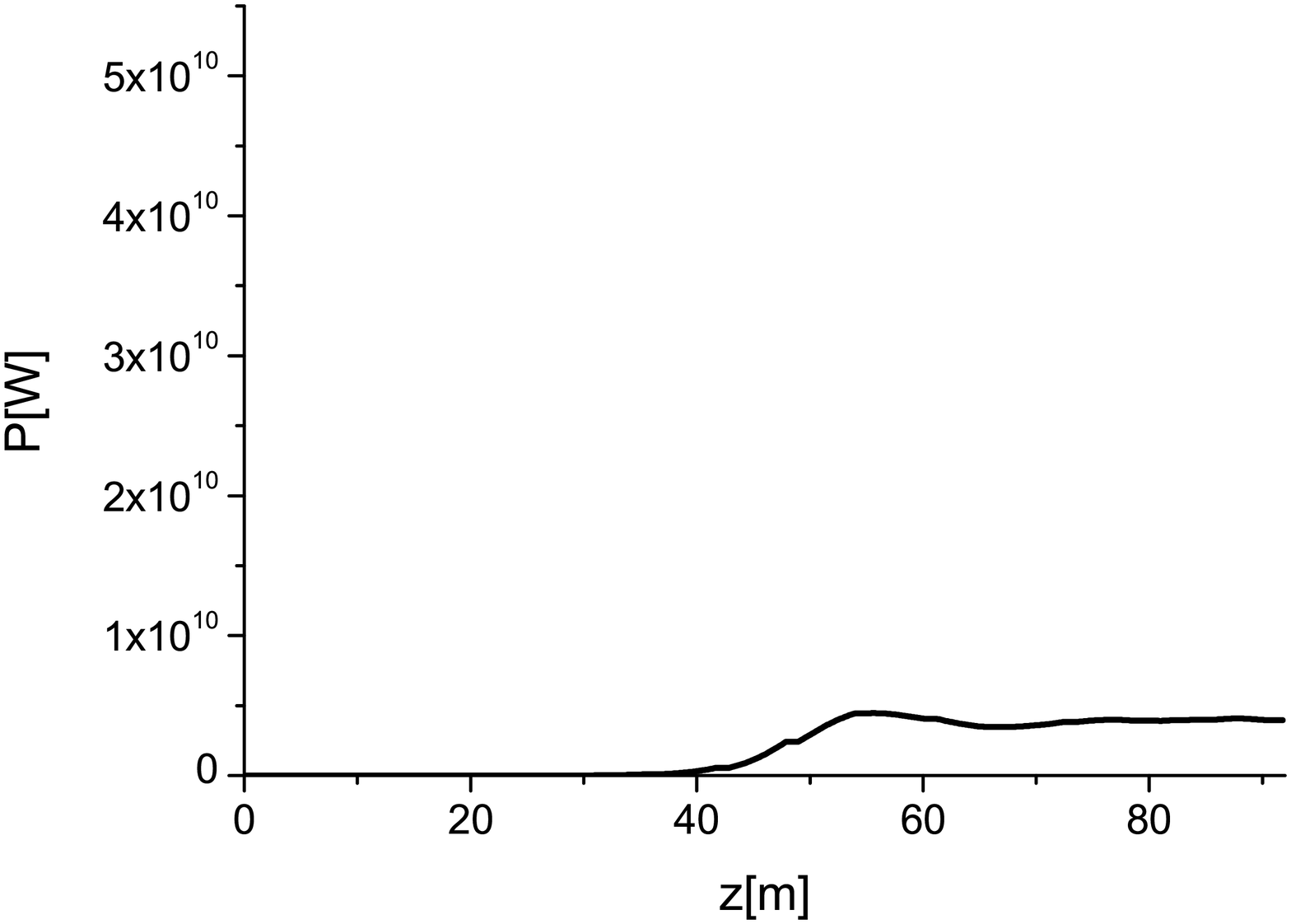}
\caption{The FEL configurations considered here refer to the SASE3
undulator line. The photon energy is 2 keV. Electron beam
characteristics are the same as  in Fig. \ref{Gcomp}. Curves are
results of numerical simulations. Left: Output power versus
undulator length. The FEL amplifier operates in the fundamental mode
in steady state nonlinear regime.  For a given electron beam energy
and undulator parameter, the value of the radiation frequency
corresponds to the maximal gain in the linear mode of operation.
Right: Output power versus undulator length. The FEL amplifier
operates in harmonic lasing mode, at the seventh harmonic. }
\label{Encomp}
\end{figure}
The left and right plot in Fig. \ref{Encomp}  respectively show the
dependence of the radiation power on the undulator length in the
saturation regime. It is clearly seen from these plots that the
saturation power of the FEL operating in the harmonic lasing mode is
about ten times lower than in the fundamental lasing mode. It is
seen from these plots that the shape of the power curves at
saturation is different. In other words, the high harmonic nonlinear
mode of operation differs significantly from that of conventional
FEL operation.  Simulations show that  in our case of interest, this
fact is mainly related to the influence of the electron beam energy
spread ($2$ MeV) on the nonlinear amplification process. The effect
of the energy spread on the harmonic lasing mode  is much stronger
than on the fundamental lasing mode. Another issue to keep in mind
is that in practical situations the gain length is comparable in
both FEL amplifier configurations. It is well-known that the
bandwidth of conventional FEL amplifier $\Delta \omega/\omega \sim
1/N_w$, where $N_w$ is number of undulator periods per gain
length\footnote{The power gain in the high-gain limit can be written
in the form : $P = A \exp(z/l_g)$, where $1/l_g$ is power growth
rate .}, $l_g$. Fig. \ref{Encomp} illustrates the fact that $N_w$ is
comparable in both configurations. However, we  estimated  from Fig.
\ref{Gcomp} that in the harmonic lasing mode the frequency bandwidth
is about an order of magnitude narrower than in the case of lasing
at the fundamental.

The reason for these combined effects on bandwidth and saturation
power can be qualitatively explained as follows. Let us consider the
interaction of an electron beam with a co-propagating
electromagnetic wave in a planar undulator in terms of the exchange
of energy between electron and field. For simplicity we consider an
electromagnetic wave, linearly polarized along the $x$ direction and
a single electron oscillating in the $x$ direction as well. The
exchange of energy is proportional to the scalar product between the
transverse velocity of the electron, which has only the
$x$-component $v_x \sim \cos(k_w z)$ (where $k_w = 2\pi/\lambda_w$
and $\lambda_w$ is the period of the planar undulator), and the
electric field of the wave, which has only the $x$-component as
well, $E_x \sim \cos[\omega (z/c - t)]$. If an effective energy
exchange between electron and field has to be provided, the scalar
product between the electric field and the electron velocity,
$\vec{E}\cdot \vec{v} = E_x  v_x \sim \cos(k_w z)
\cos[\omega(z/c-t)] \sim \cos[(k_w + \omega/c) z - \omega t)]$
should be kept nearly constant along the undulator\footnote{Here we
have neglected a rapidly oscillating term on the scale $k_w z \gg
1$, which is not resonant. }, i.e synchronism should be provided.
The rate of change of the phase $ \psi = (k_w +\omega/c)z - \omega
t$ for an electron moving along the undulator axis with velocity
$v_z(z)$ can be written as

\begin{eqnarray}
\frac{d\psi}{d z} = k_w  + \frac{\omega}{c} -\frac{\omega}{v_z(z)}~.
\label{rate}
\end{eqnarray}
In the our case of interest, the longitudinal velocity of the
electron $v_z$ is close to the speed of light, $v_z \simeq c$. Due
to the electron wiggling inside the planar undulator, the
longitudinal velocity $v_z(z) = [v^2 - v_x(z)^2]^{1/2}$ is an
oscillatory function of $z$ according to

\begin{eqnarray}
\frac{1}{v_z(z)} = \frac{1}{v} + \frac{v_x^2(z)}{2 v^3} ~,\label{Kz}
\end{eqnarray}
Since $v_x$ is proportional to $\cos(k_w z)$ and $\cos^2(k_w z) = [1
+ \cos(2 k_w z)]/2$, the function $1/v_z(z)$ oscillates as twice the
undulator period. Hence the phase oscillates as twice the undulator
period too. In order to deal with such periodical wiggling, one can
use the Anger-Jacobi expansion

\begin{eqnarray}
\exp[i \alpha \sin(\theta)] = \sum_{m=-\infty}^{\infty} J_m(\alpha)
\exp[i m \theta]\label{AJ}
\end{eqnarray}

Using this relation we obtain

\begin{eqnarray}
\exp(i \psi) \sim \sum_m A_m \exp\left[i \left(k_w +
\frac{\omega}{c} - \frac{\omega}{\langle v(z) \rangle} +2 m
k_w\right)z \right]\label{cospsi}
\end{eqnarray}
where

\begin{eqnarray}
\frac{1}{\langle v(z) \rangle} = \frac{1}{c} + \frac{1+K^2/2}{2 c
\gamma^2} ~,\label{vz}
\end{eqnarray}
where $\gamma$ is the relativistic Lorentz factor, and $K$ is the
dimensionless undulator parameter, related to the undulator period
$\lambda_w$ and to the undulator peak magnetic field $H$ as $K =
0.934 \cdot \lambda_w [\mathrm{cm}]\cdot H[\mathrm{T}]$.  The energy
change of the electron along the undulator is proportional to $\int
\vec{E} \cdot \vec{v} dz$. The integrand does not contribute
appreciably unless the arguments in the exponential, oscillating
functions vanish. In other words, resonance condition requires the
phase to be independent of z, and one obtains for $m = 0,1, ..$:

\begin{eqnarray}
(2m + 1) k_w + \frac{\omega}{c} - \frac{\omega}{\langle
v_z(z)\rangle} = 0 ~.\label{reso}
\end{eqnarray}
Let us now consider the two cases in Fig. \ref{compro}(a) and Fig.
\ref{compro}(b). In the first case, the target frequency coincides
with the fundamental $\omega_0 = \omega_1$, and Eq. (\ref{reso})
simply becomes

\begin{eqnarray}
&& \left(\frac{d  \psi}{d z} \right)_1=   \frac{\omega_0
(1+K_1^2/2)}{2 c \gamma^2} - k_w = 0~,\label{reso1}
\end{eqnarray}
where $\gamma$ is the relativistic Lorentz factor. In the second
case, the target frequency coincides with the nth harmonic of the
fundamental $\omega_0 = n \omega_1$, with $n= 2m+1$,  so that Eq.
(\ref{reso}) becomes

\begin{eqnarray}
&& \left( \frac{d \psi }{d z} \right)_n=  \frac{\omega_0
(1+K_n^2/2)}{2 c \gamma^2} - n k_w = 0~.\label{reson}
\end{eqnarray}
By inspecting Eq. (\ref{reso1}) and Eq. (\ref{reson}) it is
straightforward to see that the quantity ${\omega_0 (1+K_n^2/2)}/{(2
c \gamma^2)}$ in Eq. (\ref{reson}) must be $n$ times larger than the
quantity ${\omega_0 (1+K_1^2/2)}/{(2 c \gamma^2)}$ in Eq.
(\ref{reso1}), for the resonance condition to hold. In particular,
in order to produce the same rate of phase change $d \psi/dz$ one
needs, for example, a deviation from $\omega_0$ or a deviation from
$\gamma$ that is $n$ times smaller in Eq. (\ref{reson}), compared to
Eq. (\ref{reso1}). Let us now come back to our statement that the
gain lengths in our two FEL amplifier configurations are comparable.
It should be clear that the gain strongly depends on the rate of
phase change, that is the detuning from resonance $d \psi/dz$. It is
important to realize that, similarly as the gain lengths, also the
detuning bandwidths are comparable in the two FEL amplifier
configurations under study.  One qualitatively concludes that there
must be a significant difference in frequency bandwidth and
saturation power for the harmonic lasing mode compared to the
fundamental lasing mode of FEL operation, as exemplified in Fig.
\ref{Gcomp} and Fig. \ref{Encomp}.

\begin{figure}
\includegraphics[width=0.50\textwidth]{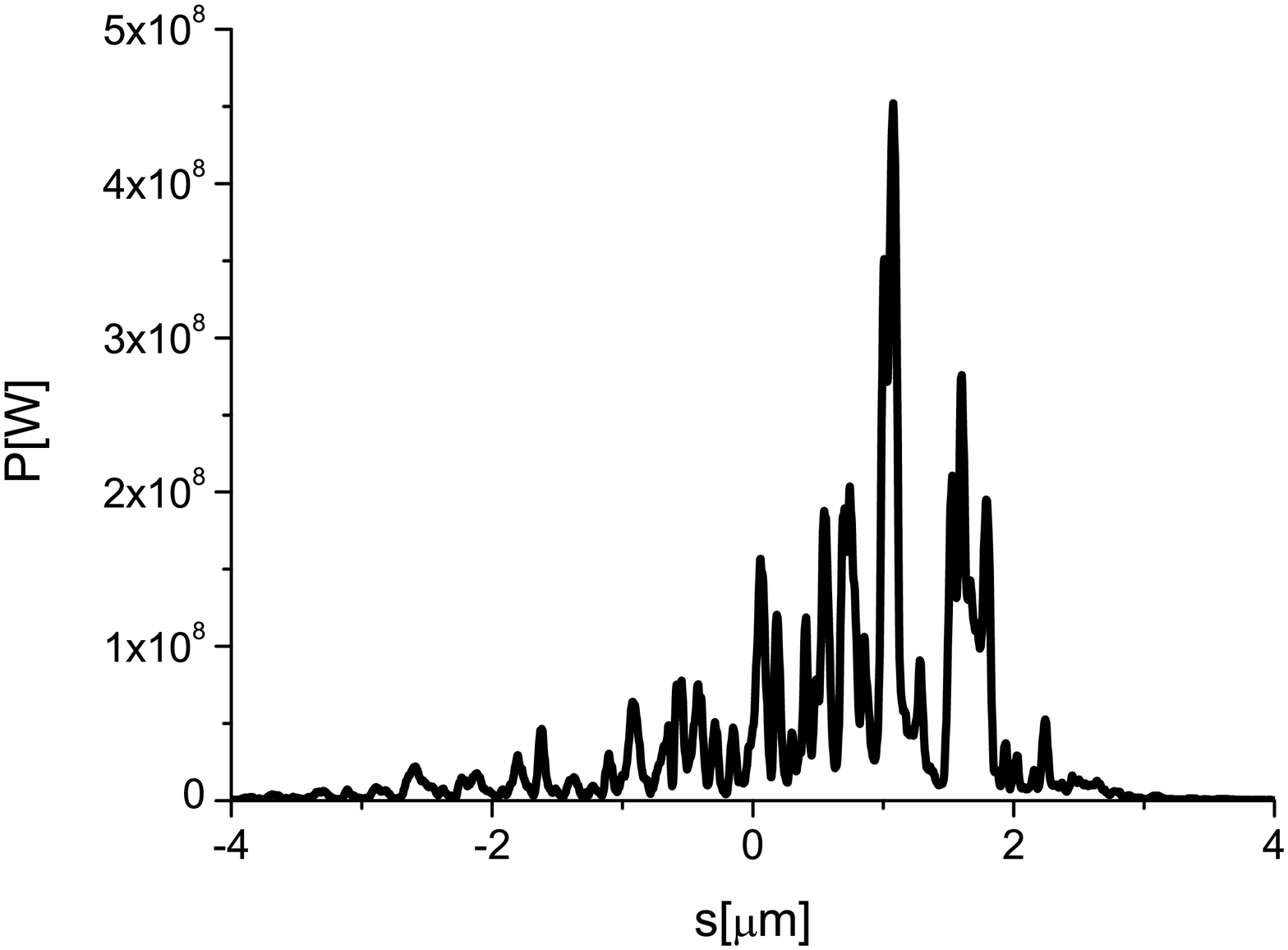}
\includegraphics[width=0.50\textwidth]{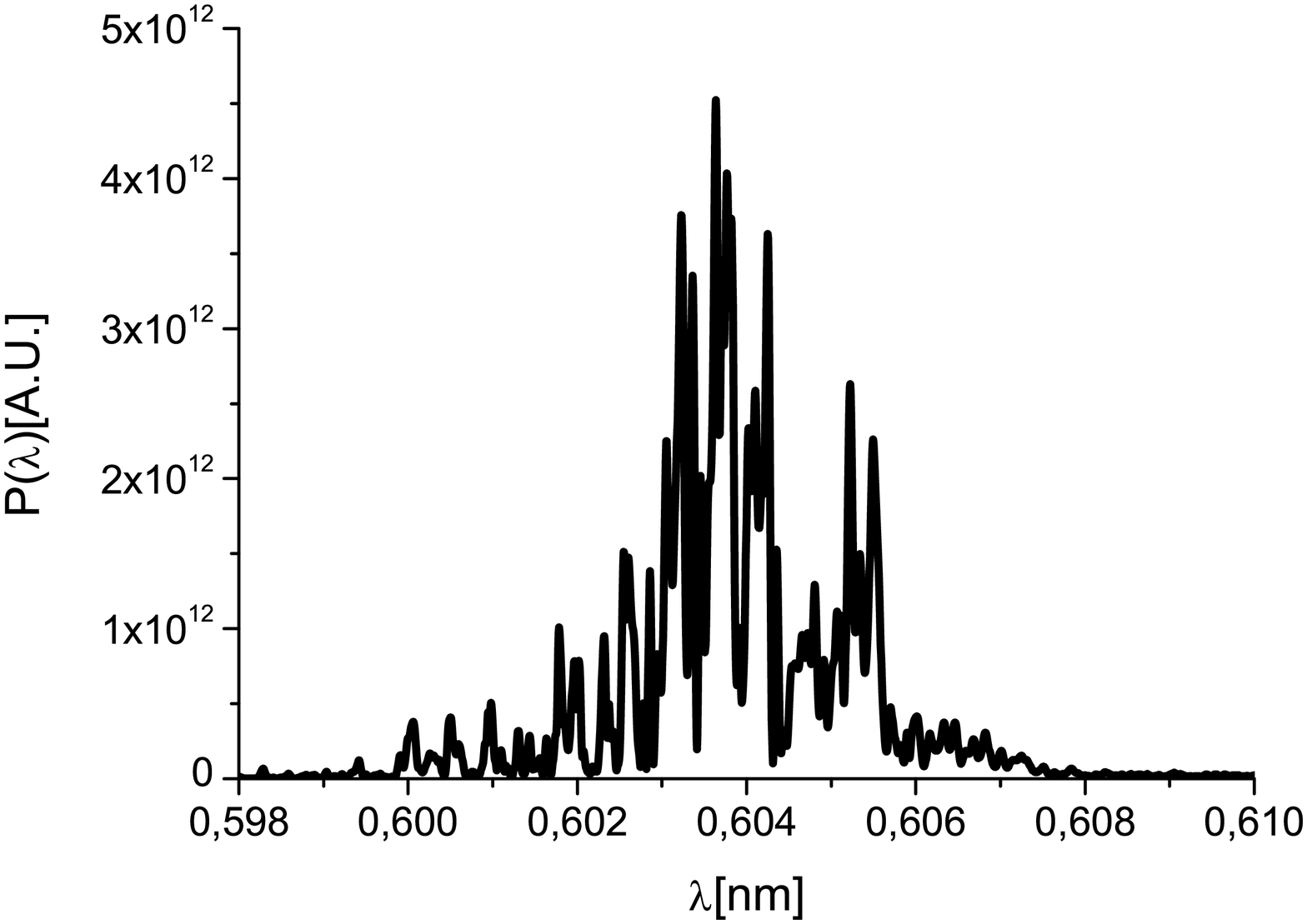}
\caption{SASE radiation power and spectrum at the exit of the first
SASE3 undulator part U1 with $5$ cells resonant at $0.6$ nm. Results
obtained with Genesis code using an electron beam with 0.1 nC charge
and 10.5 GeV energy (see section 3 for details). } \label{PSPin1}
\end{figure}
\begin{figure}
\includegraphics[width=0.50\textwidth]{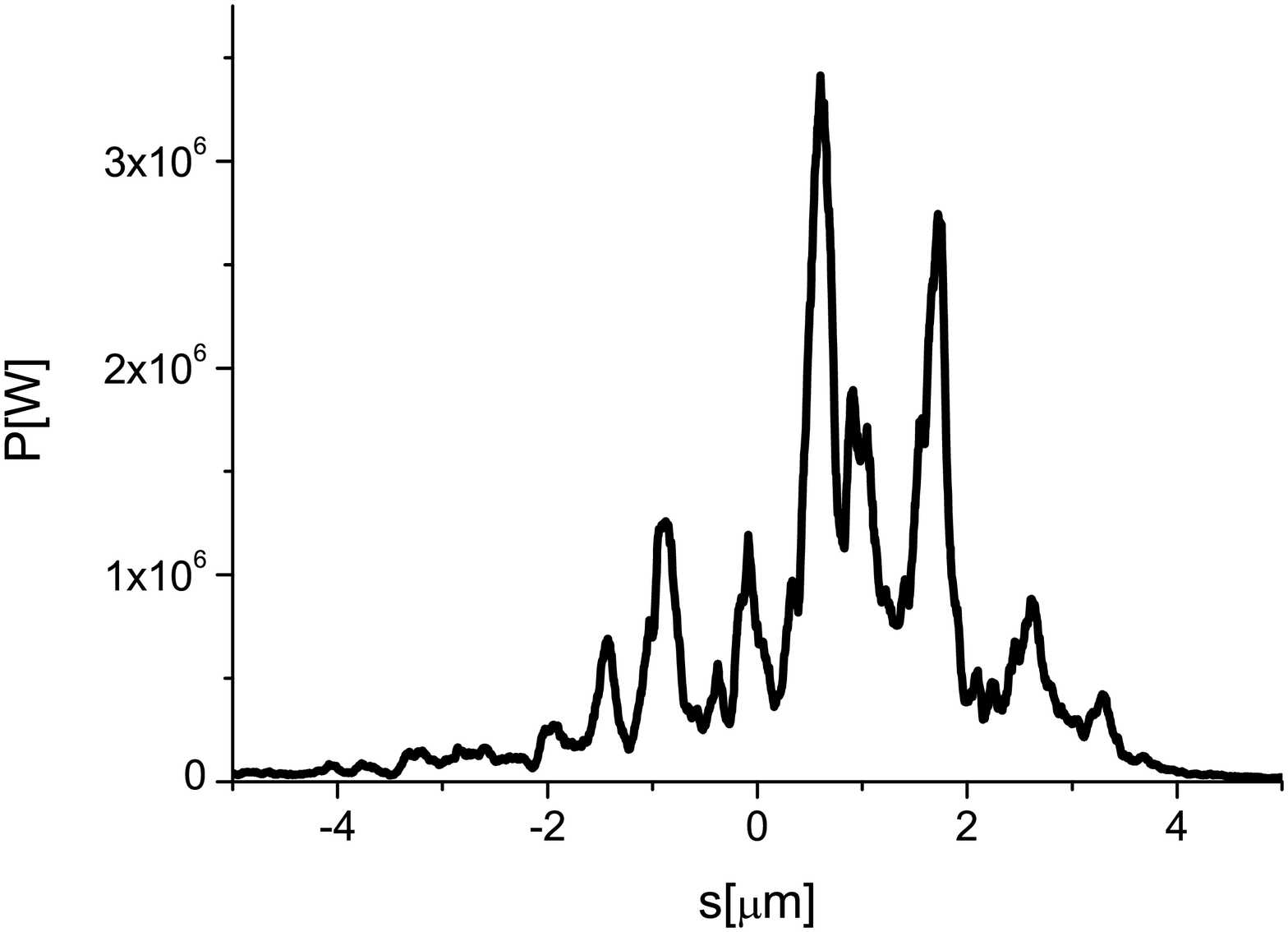}
\includegraphics[width=0.50\textwidth]{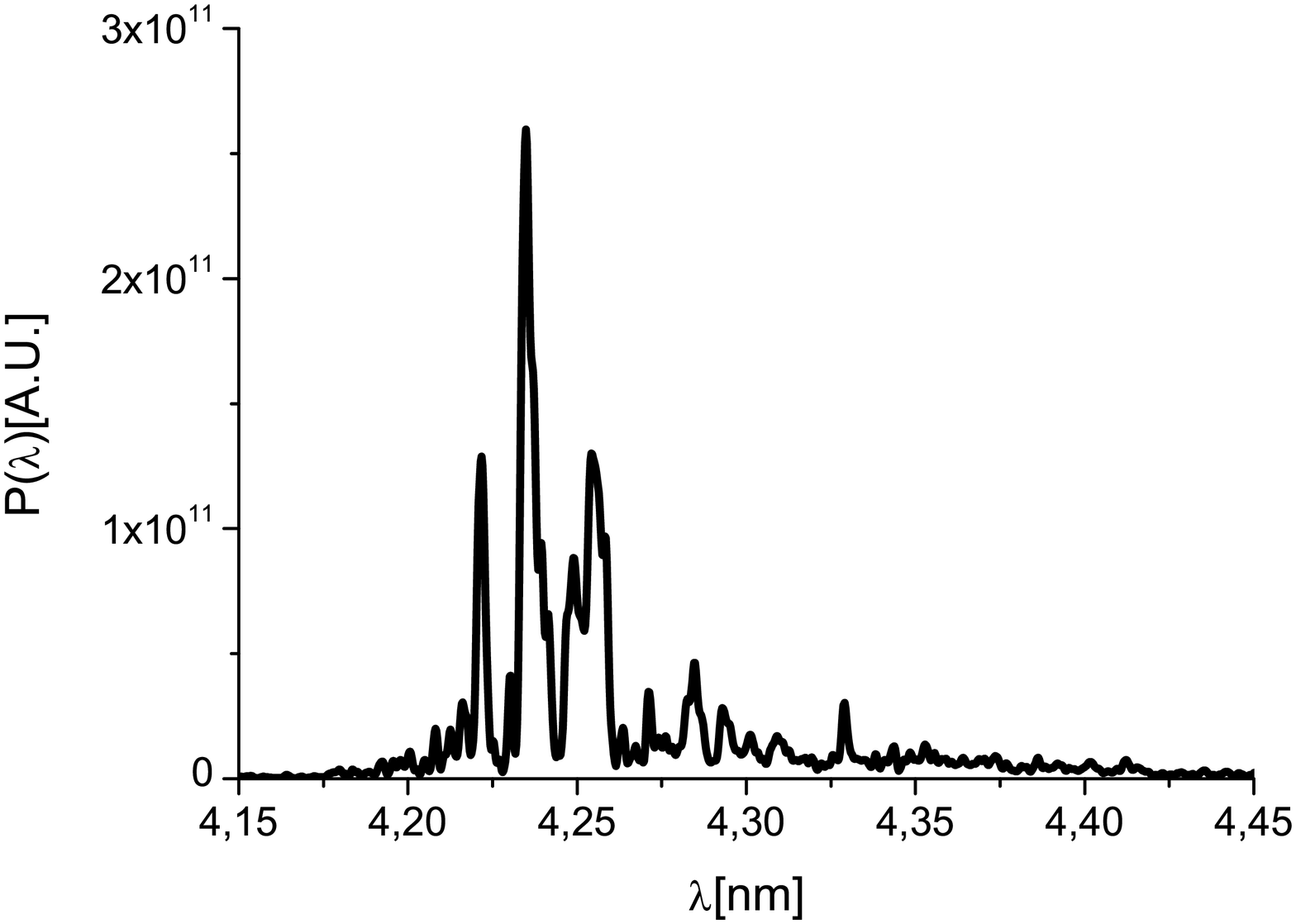}
\includegraphics[width=0.50\textwidth]{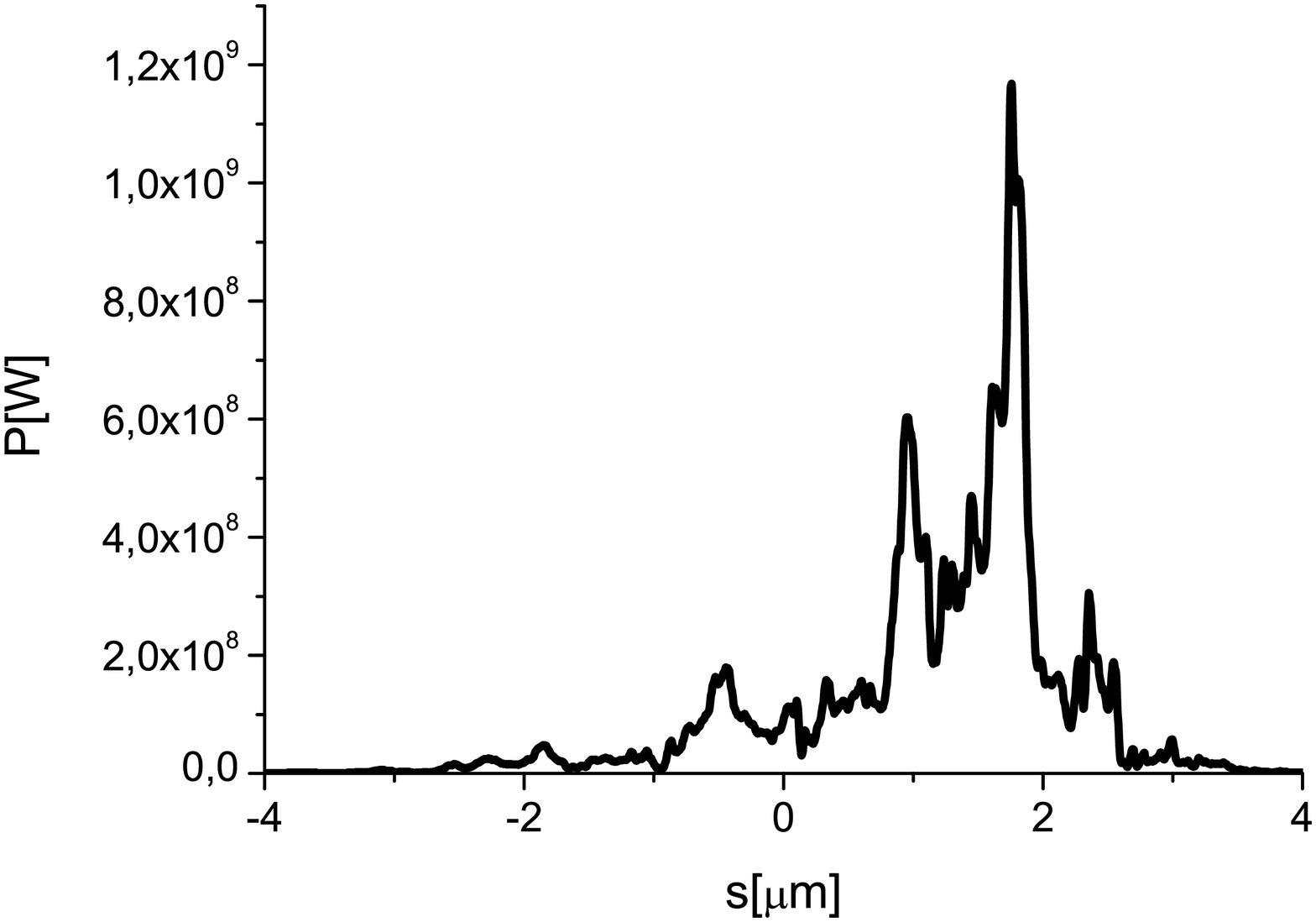}
\includegraphics[width=0.50\textwidth]{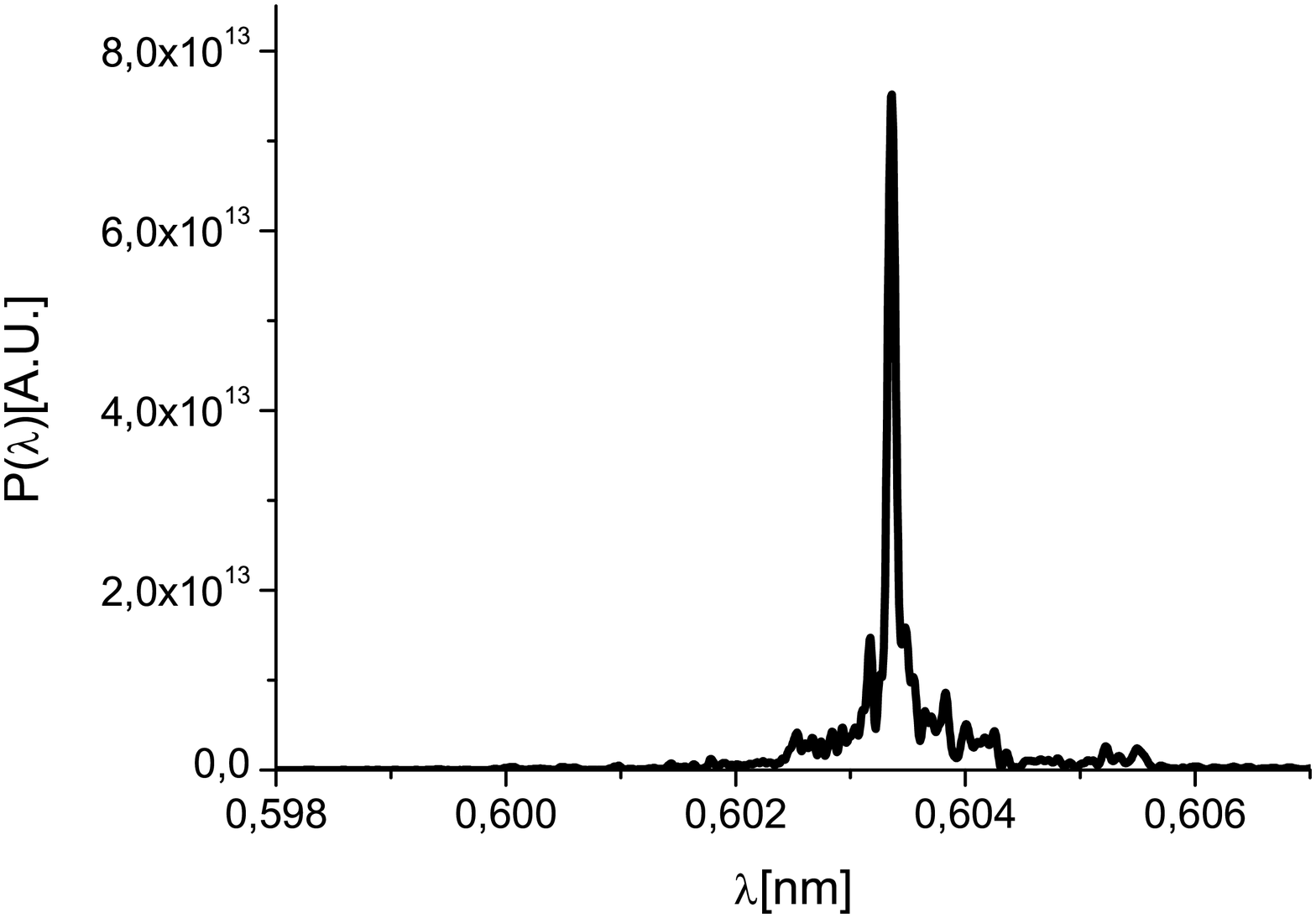}
\caption{SASE radiation power and spectrum at the exit of the second
undulator part U2 (slippage-boosted section). The SASE radiation
generated in U1 is purified in U2, which consists of $2$ cells
resonant at $4.2$ nm. The fundamental radiation at $4.2$ nm is
seeded by shot noise. The harmonic radiation is seeded by that
produced in U1. Top row: Results of numerical simulations for the
radiation at the fundamental produced in U2. Bottom row: Results of
numerical simulations for harmonic radiation amplified in U2. }
\label{PSPU21}
\end{figure}
Results of our simulations show that in order to operate in the deep
linear regime, the number of cells in the undulator U1 should be
equal to five. The output power and spectrum after the first
undulator tuned to $0.6$ nm is shown in Fig. \ref{PSPin1} for a
single shot realization. The length of the slippage-boosted section
U2 should be properly chosen to make sure that the FEL power at the
fundamental wavelength is much lower than that at the chosen
harmonic. This is possible because radiation at the fundamental
wavelength starts from shot noise, while radiation at the target
wavelength is seeded by the radiation produced in U1, and the seed
power is much higher than the effective shot noise power. The output
power and spectrum of fundamental and harmonic radiation pulse after
the U2 undulator tuned to $4.2$ nm are shown in the left and right
plot of Fig. \ref{PSPU21}. Since the FEL power at the fundamental
wavelength of $4.2$ nm, which is about $1$ MW, is much lower than
that at $0.6$ nm, which is about $1$ GW, phase shifters are not
needed to suppress the lasing at fundamental harmonic.

\begin{figure}
\includegraphics[width=0.50\textwidth]{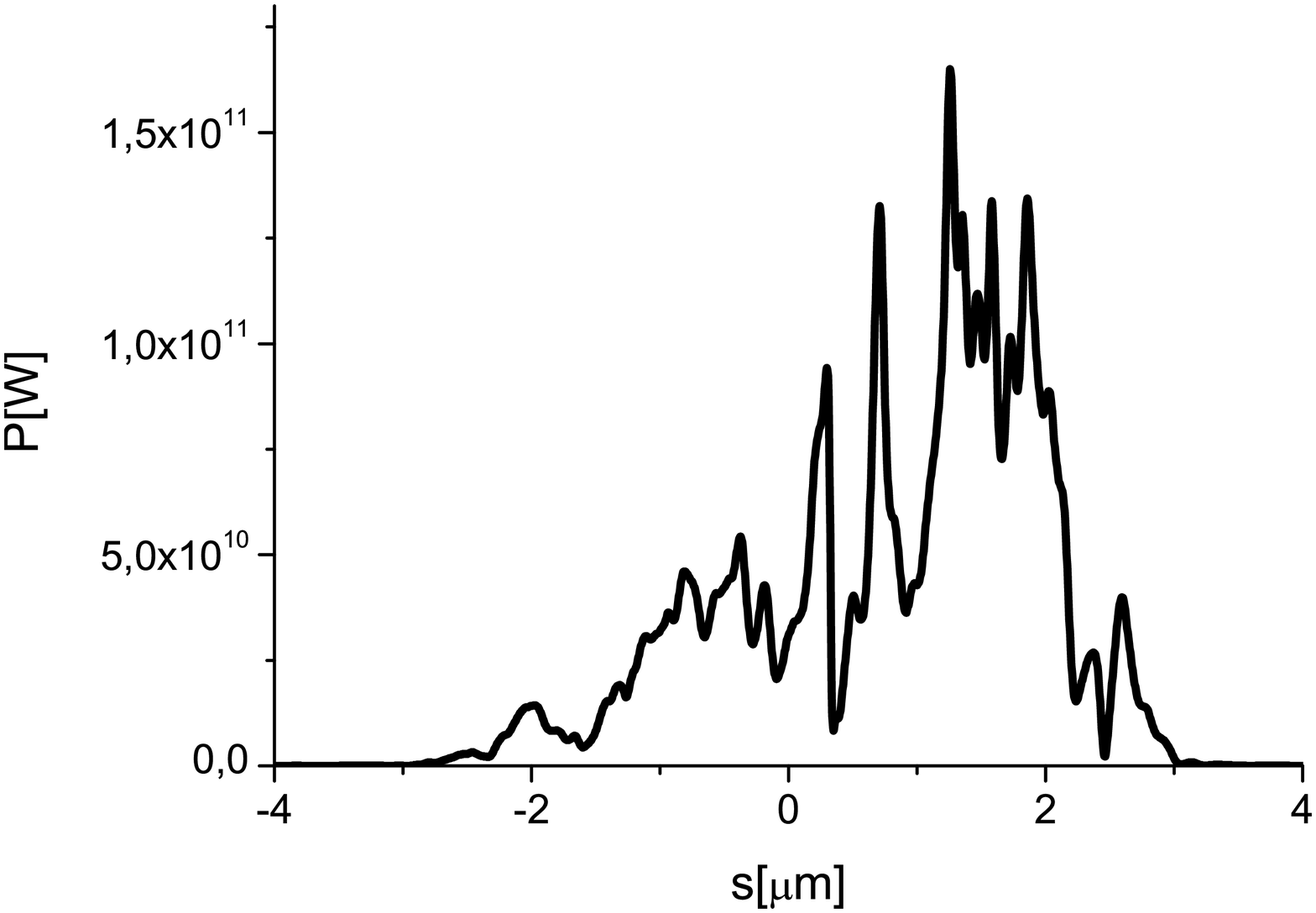}
\includegraphics[width=0.50\textwidth]{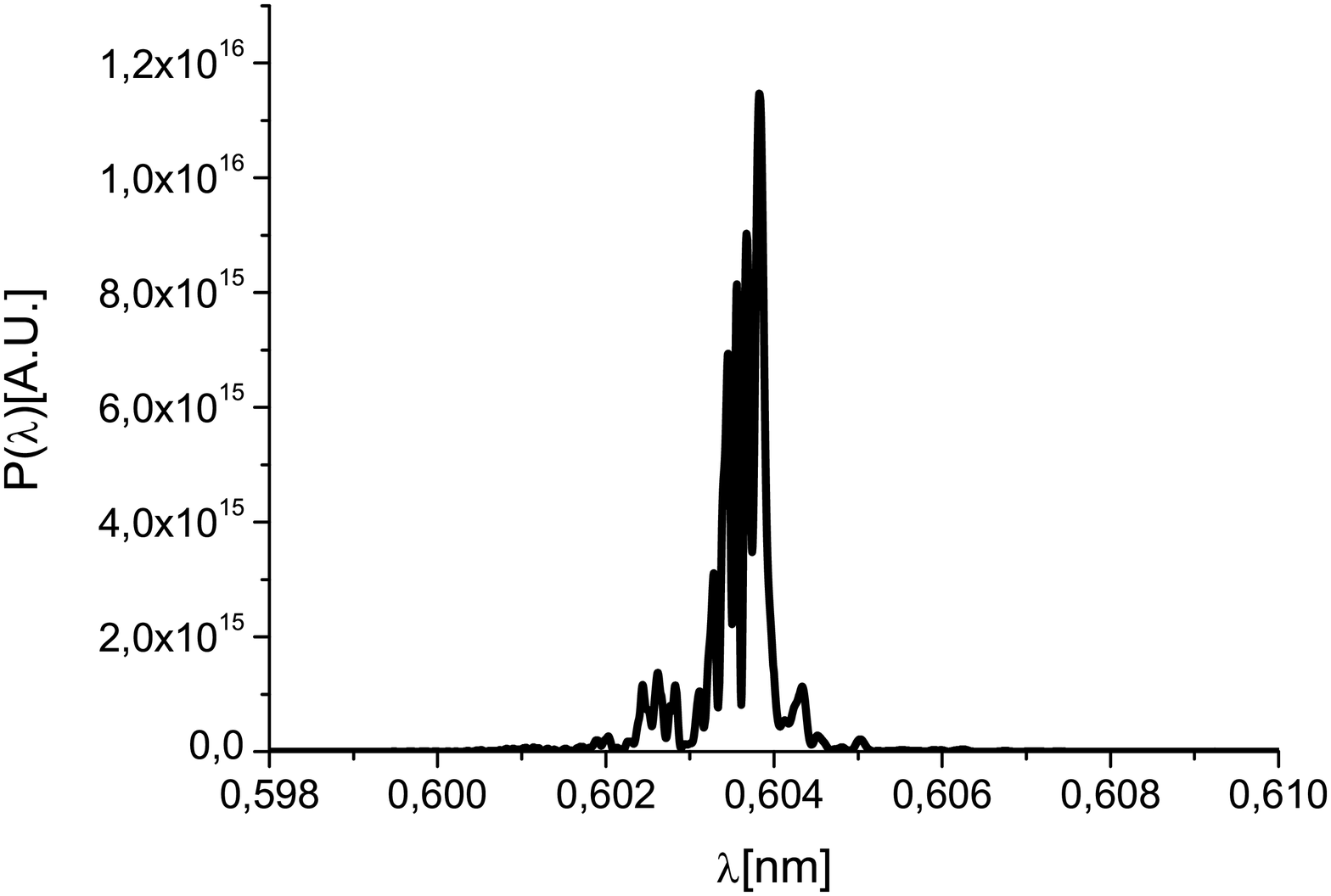}
\includegraphics[width=0.50\textwidth]{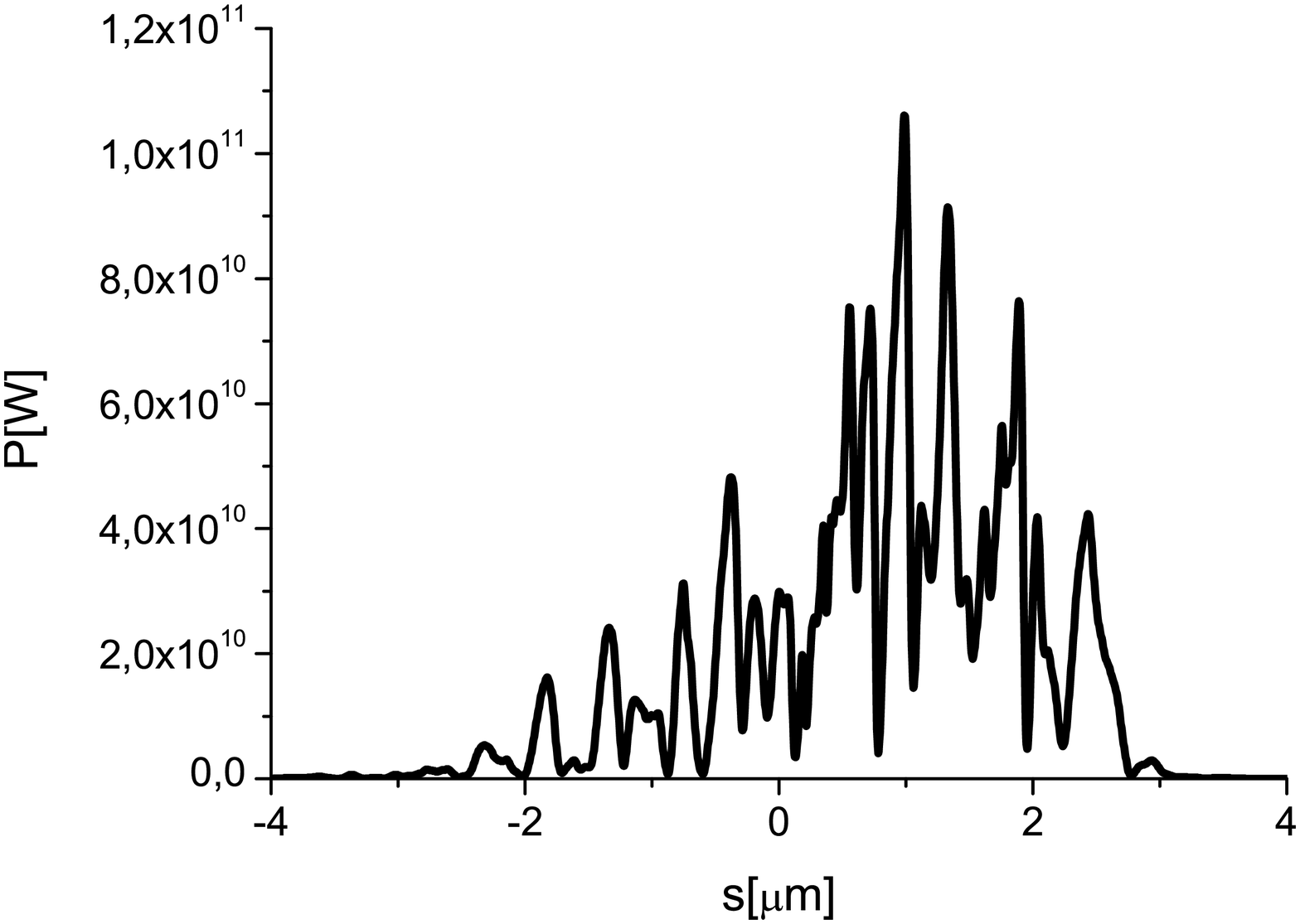}
\includegraphics[width=0.50\textwidth]{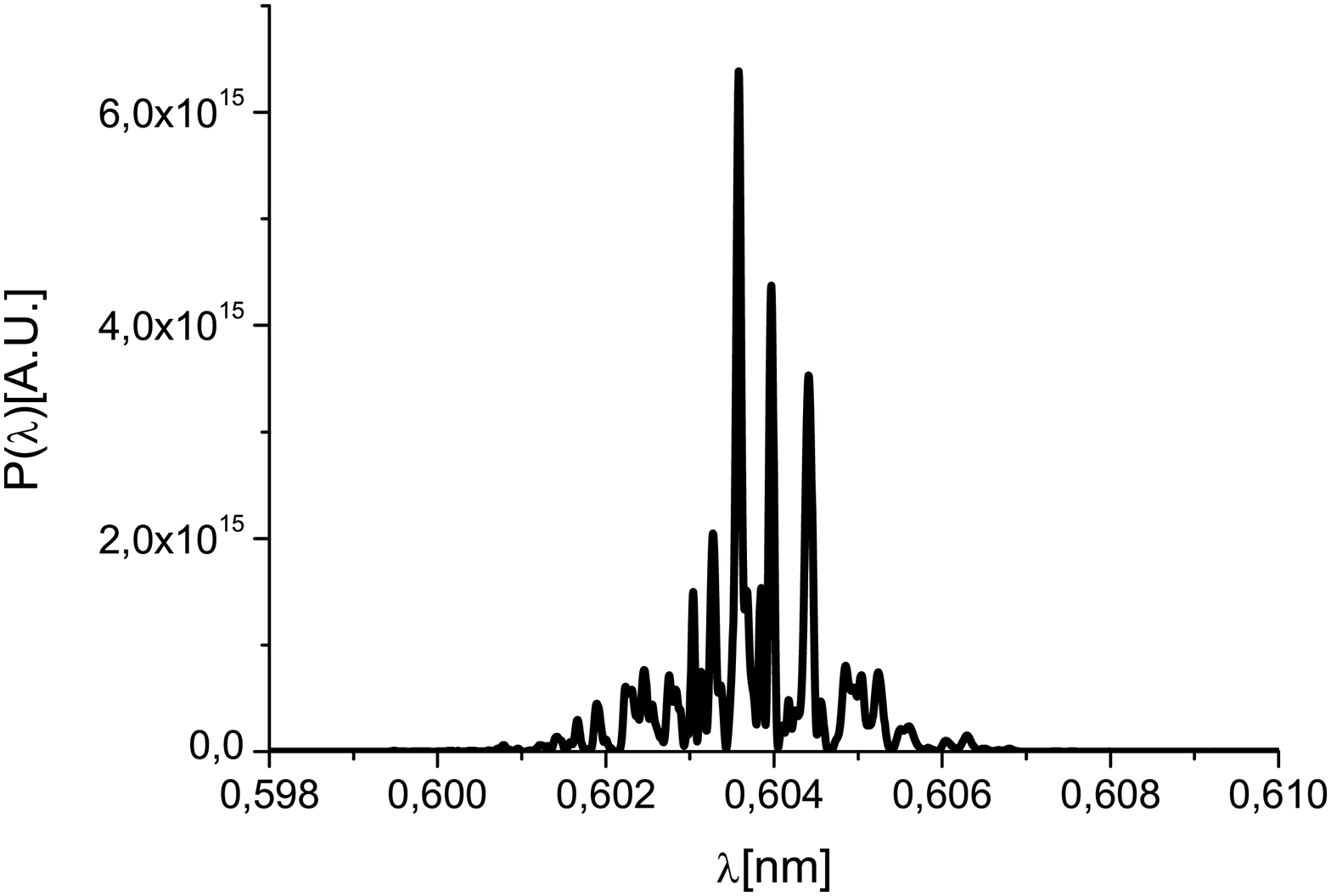}
\caption{Power and spectrum produced in the pSASE mode (top row) and
in the standard SASE mode (bottom row) at saturation without
undulator tapering. Scales of spectral energy density are the same
for both cases. } \label{PSPoutsat1}
\end{figure}

\begin{figure}
\includegraphics[width=0.50\textwidth]{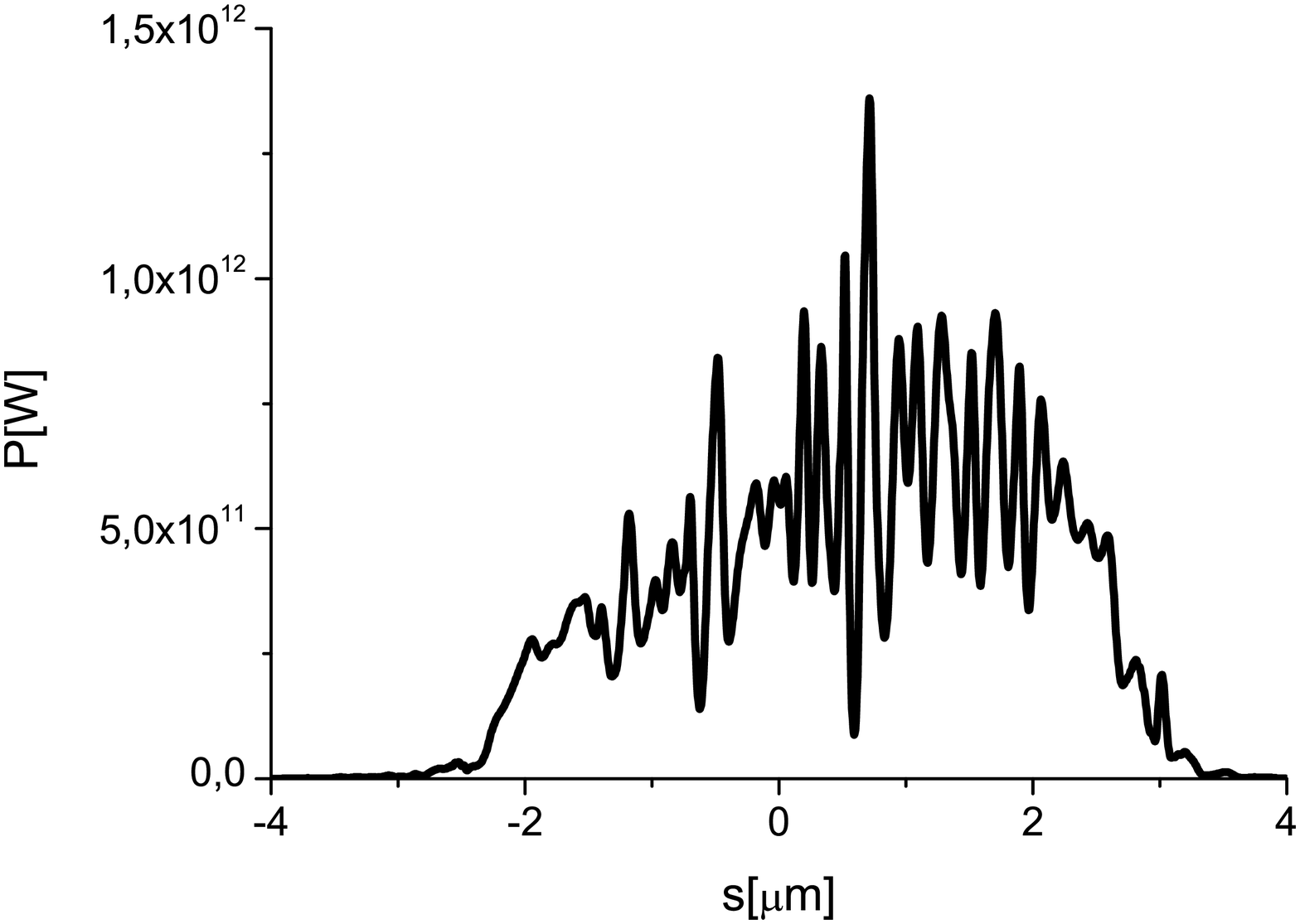}
\includegraphics[width=0.50\textwidth]{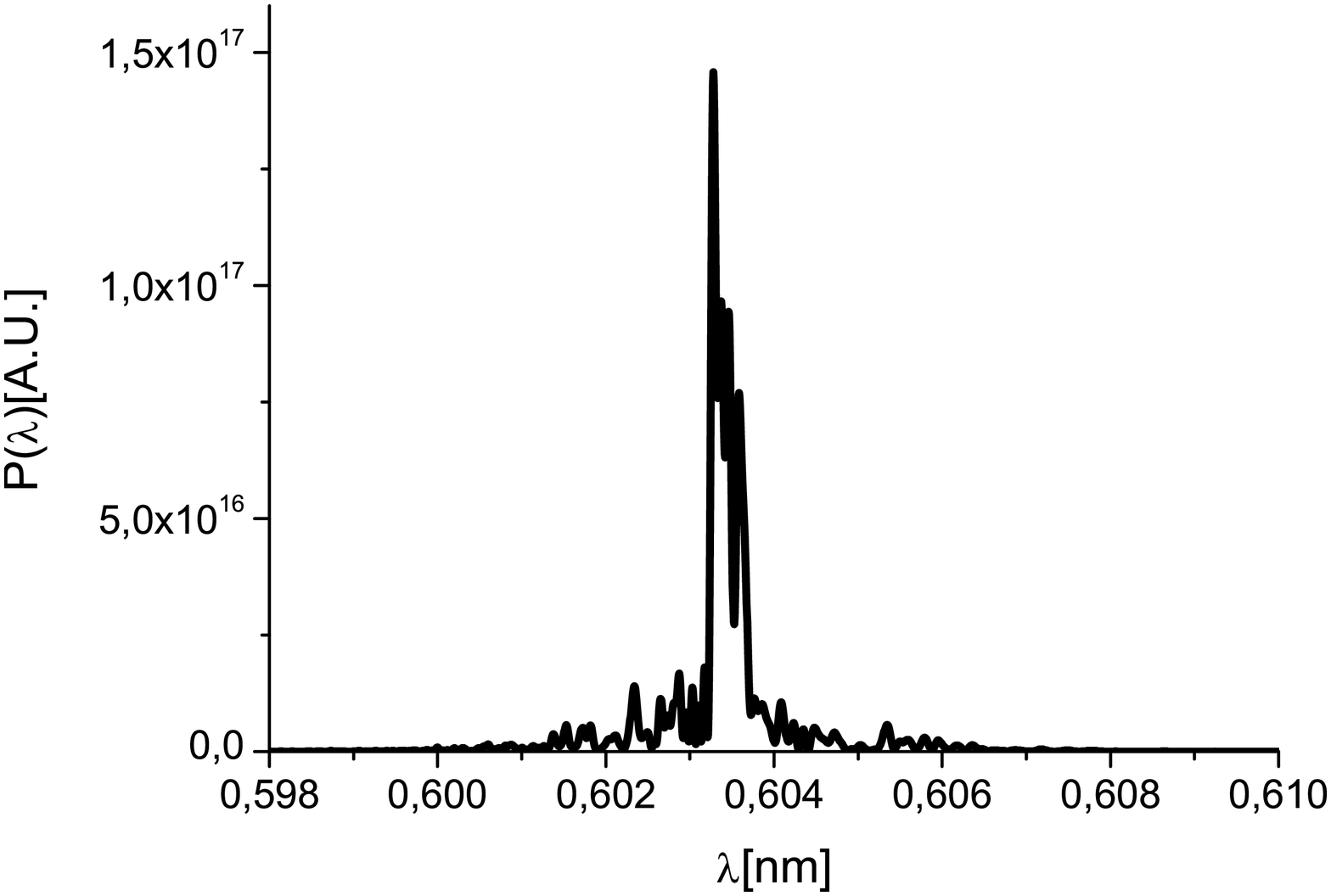}
\caption{Power and spectrum produced at the exit of the pSASE setup
with undulator tapering. The scale of the spectral energy density is
the same as Fig. \ref{PSPoutsat1}. } \label{PSPoutap1}
\end{figure}

\begin{figure}
\begin{center}
\includegraphics[width=0.50\textwidth]{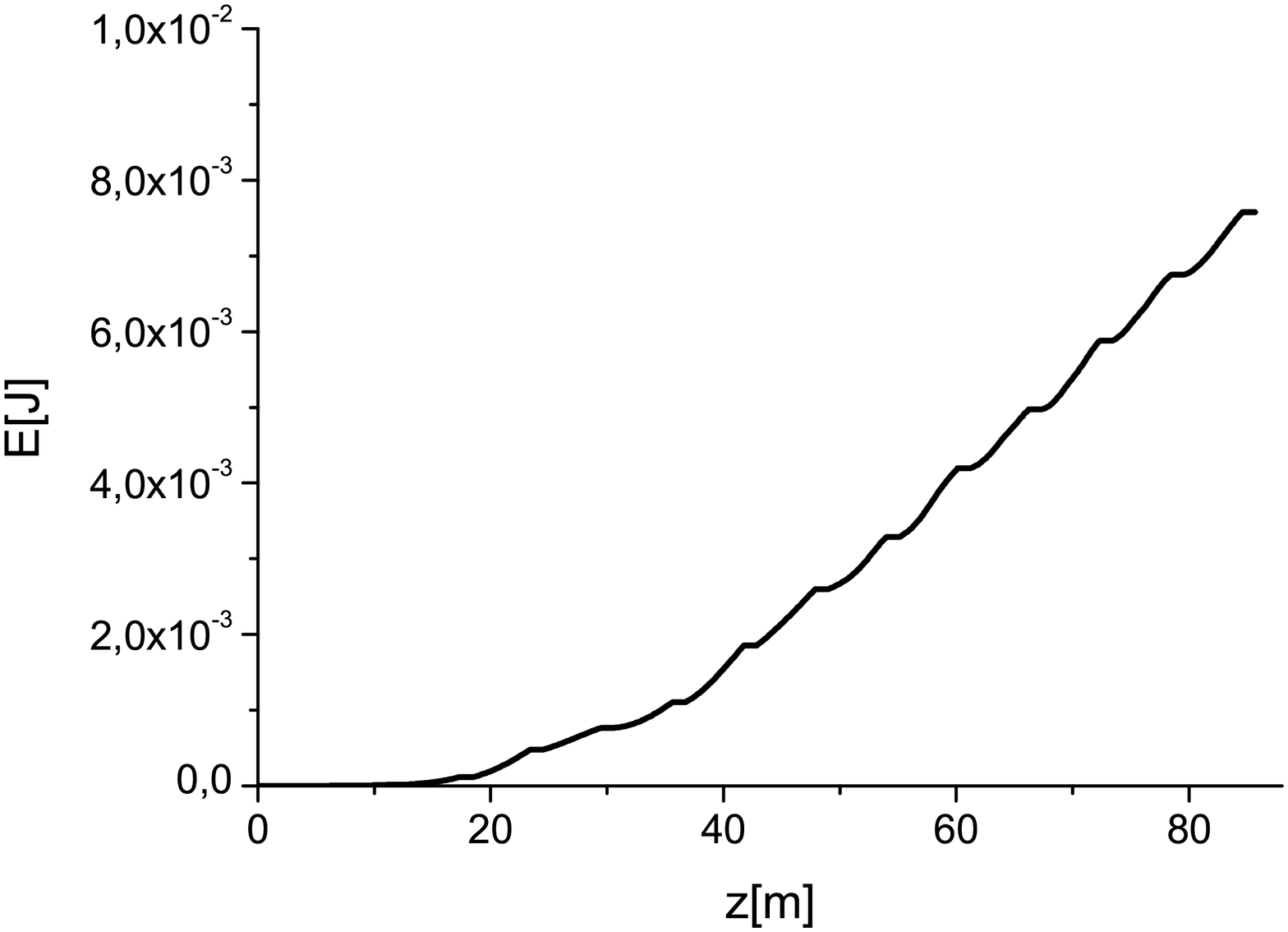}
\end{center}
\caption{Evolution of the output energy in the photon pulse as a
function of the distance inside the output undulator U3 in the pSASE
mode with tapering.} \label{Enout1}
\end{figure}
The output undulator U3 consists of two sections. The first  section
is composed by an uniform undulator, the second section by a tapered
undulator. The purified pulse is exponentially amplified passing
through the first uniform part of the output undulator. This section
is long enough, $5$ cells, in order to reach saturation, which
yields about 50 GW power Fig. \ref{PSPoutsat1} (top row).  The
radiation power profile and spectra for the SASE3 undulator beamline
working in the nominal SASE mode is shown for comparison in Fig.
\ref{PSPoutsat1} (bottom row). The power level for both modes of
operation are similar, but the spectral density for the pSASE case
is significantly higher than for the nominal SASE case. Finally, in
the second part of the output undulator U3 the purified FEL output
is enhanced up to about $0.6$ TW taking advantage of a taper of the
undulator magnetic field over the last $9$ cells after saturation.
The output power and spectrum of the entire setup is shown in Fig.
\ref{PSPoutap1}. The evolution of the output energy in the photon
pulse as a function of the distance inside the output undulator is
reported in Fig. \ref{Enout1}. As one can see, the photon spectral
density for the output TW-level pulse is about $30$ times higher
than that for the nominal SASE pulse at saturation.

\section{\label{sec:study} FEL studies}

In this section we present a more thorough feasibility study of the
pSASE setup described in the previous section with the help of the
FEL code Genesis 1.3 \cite{GENE} running on a parallel machine.
Results are presented for the SASE3 FEL line of the European XFEL,
based on a statistical analysis consisting of $100$ runs. The
overall beam parameters used in the simulations are presented in
Table \ref{tt1}.

\begin{table}
\caption{Parameters for the mode of operation at the European XFEL
used in this paper.}

\begin{small}\begin{tabular}{ l c c}
\hline & ~ Units &  ~ \\ \hline
Undulator period      & mm                  & 68     \\
Periods per cell      & -                   & 73   \\
Total number of cells & -                   & 21    \\
Intersection length   & m                   & 1.1   \\
Energy                & GeV                 & 10.5 \\
Charge                & nC                  & 0.1\\
\hline
\end{tabular}\end{small}
\label{tt1}
\end{table}

\begin{figure}[tb]
\includegraphics[width=0.5\textwidth]{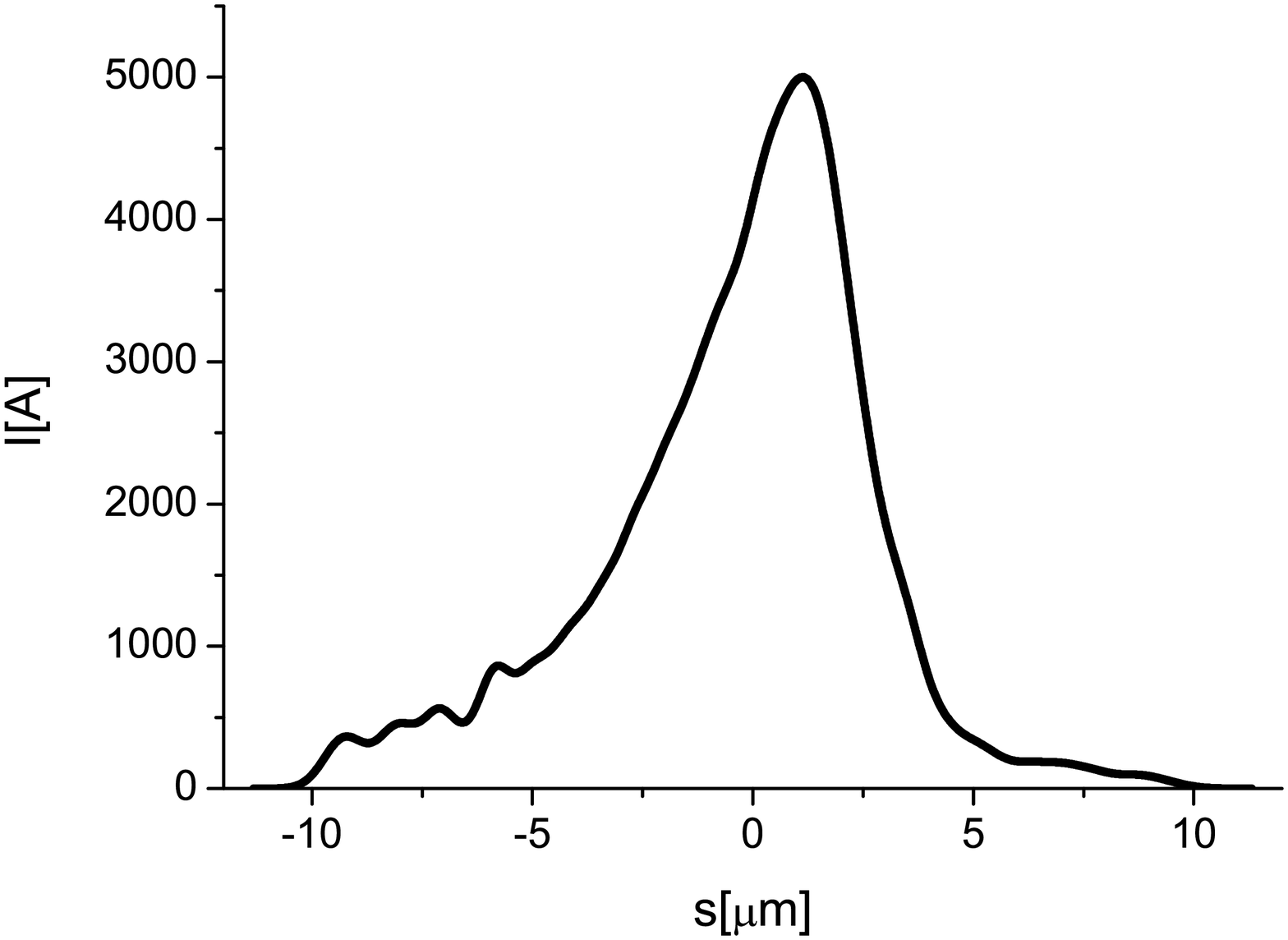}
\includegraphics[width=0.5\textwidth]{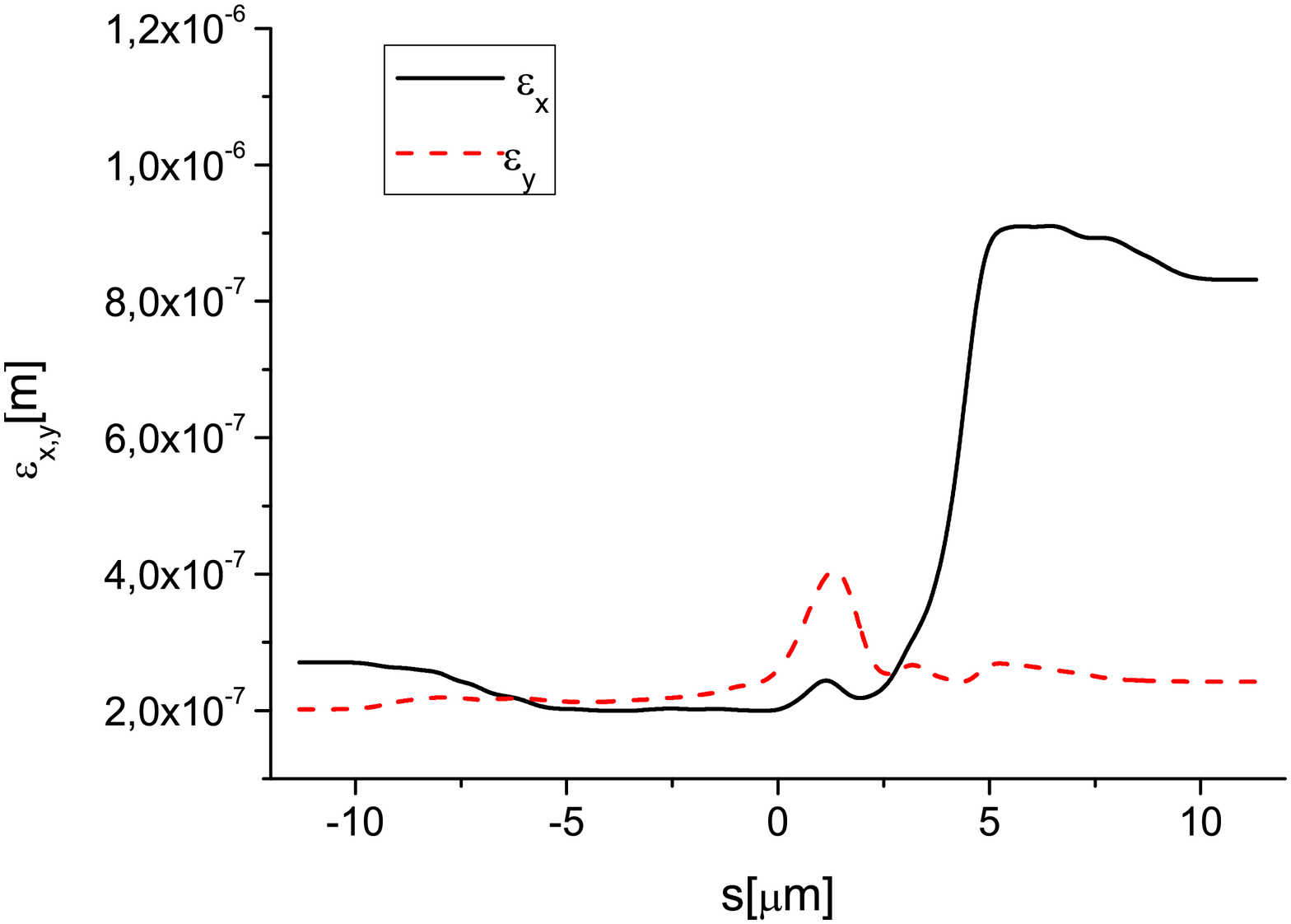}
\includegraphics[width=0.5\textwidth]{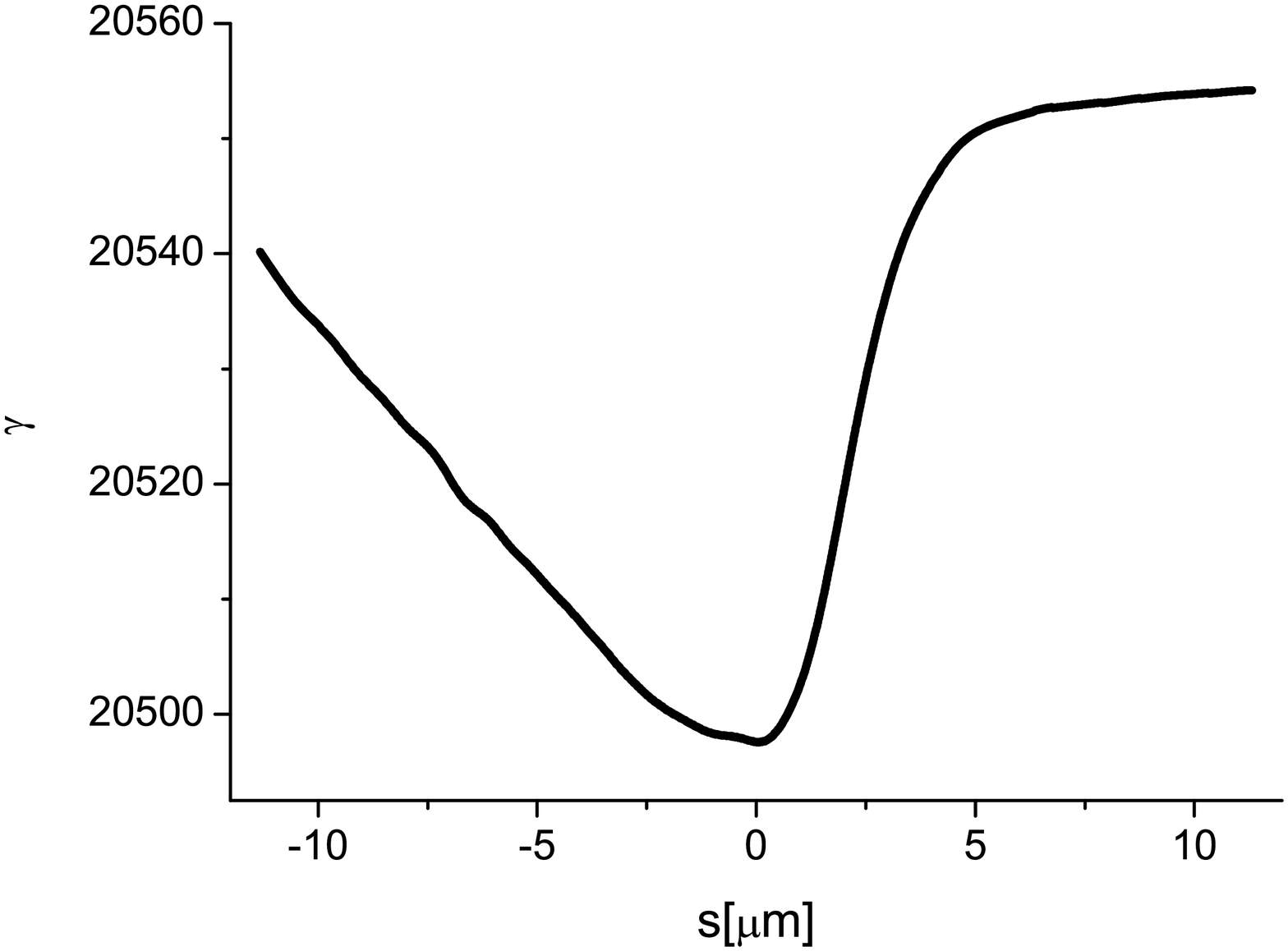}
\includegraphics[width=0.5\textwidth]{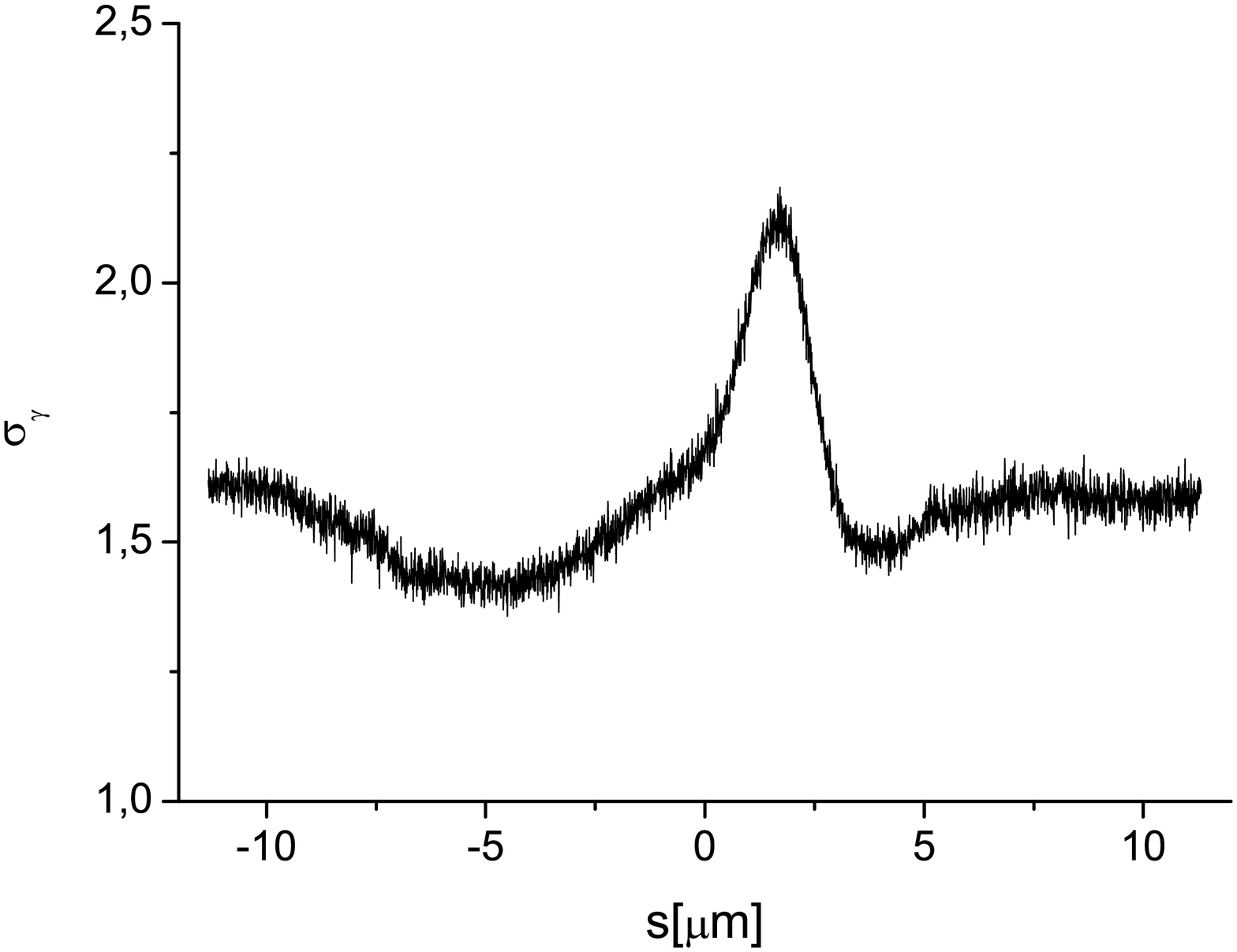}
\begin{center}
\includegraphics[width=0.5\textwidth]{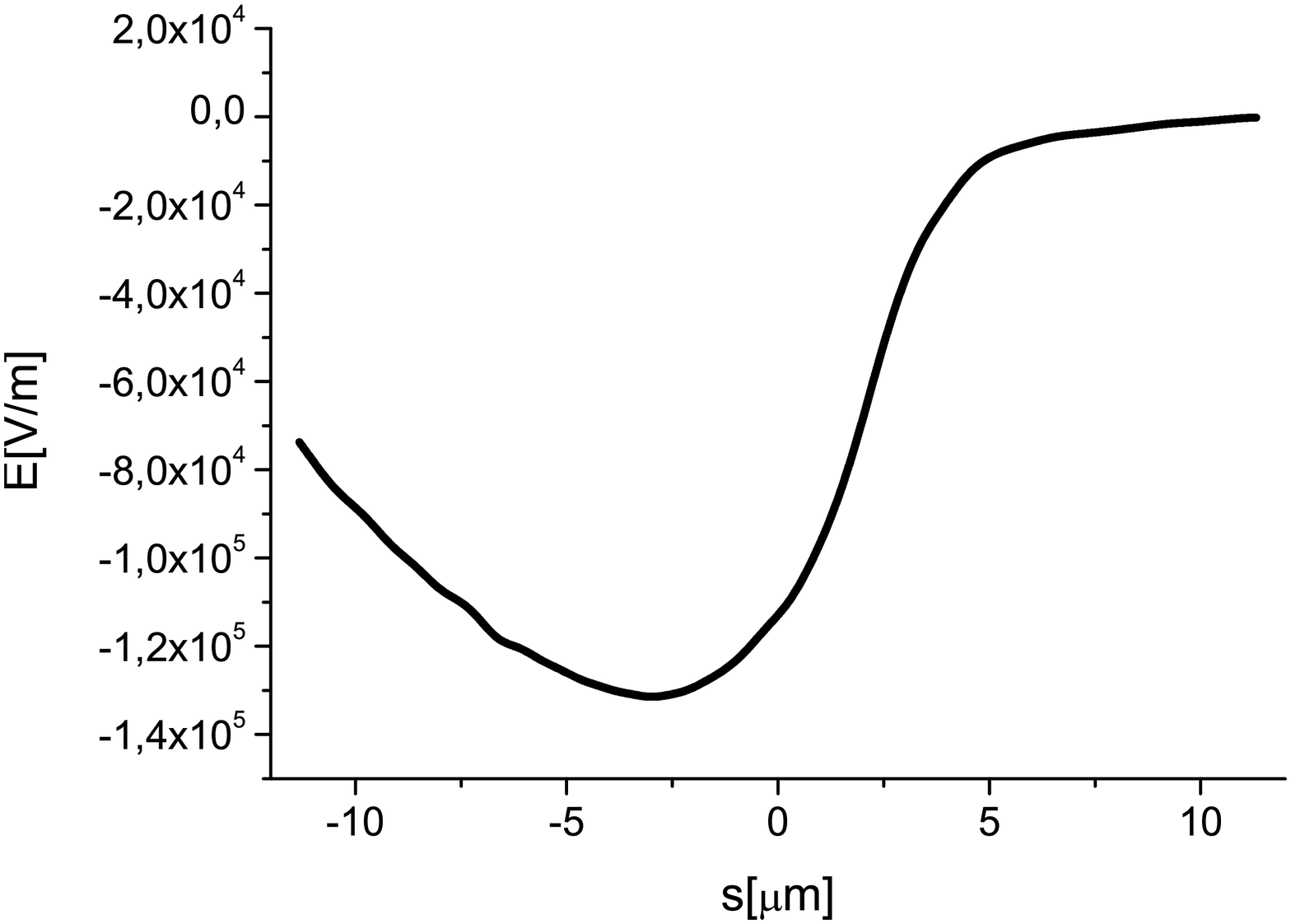}
\end{center}
\caption{Results from electron beam start-to-end simulations at the
entrance of SASE3. (First Row, Left) Current profile. (First Row,
Right) Normalized emittance as a function of the position inside the
electron beam. (Second Row, Left) Energy profile along the beam.
(Second Row, Right) Electron beam energy spread profile. (Bottom
row) Resistive wakefields in the SASE3 undulator.} \label{s2E}
\end{figure}
\begin{figure}[tb]
\begin{center}
\includegraphics[width=0.5\textwidth]{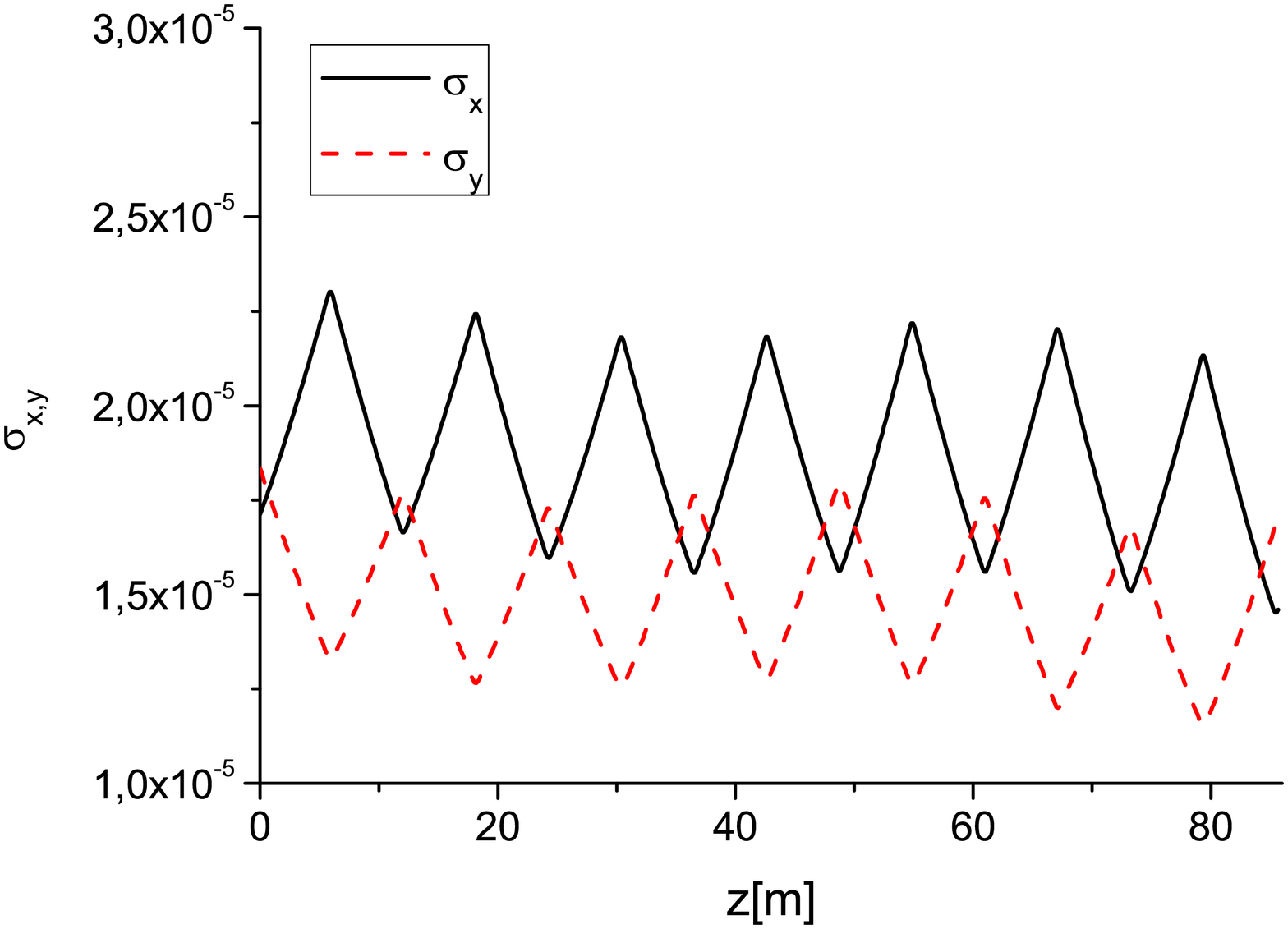}
\end{center}
\caption{Evolution of the horizontal and vertical dimensions of the
electron bunch as a function of the distance inside the SASE3
undulator. The plots refer to the longitudinal position inside the
bunch corresponding to the maximum current value.} \label{sigma}
\end{figure}
The nominal beam parameters at the entrance of the SASE3 undulator,
and the resistive wake inside the undulator are shown in Fig.
\ref{s2E}, \cite{ZAGO}. The evolution of the transverse electron
bunch dimensions is plotted in Fig. \ref{sigma}.

\begin{figure}
\includegraphics[width=0.50\textwidth]{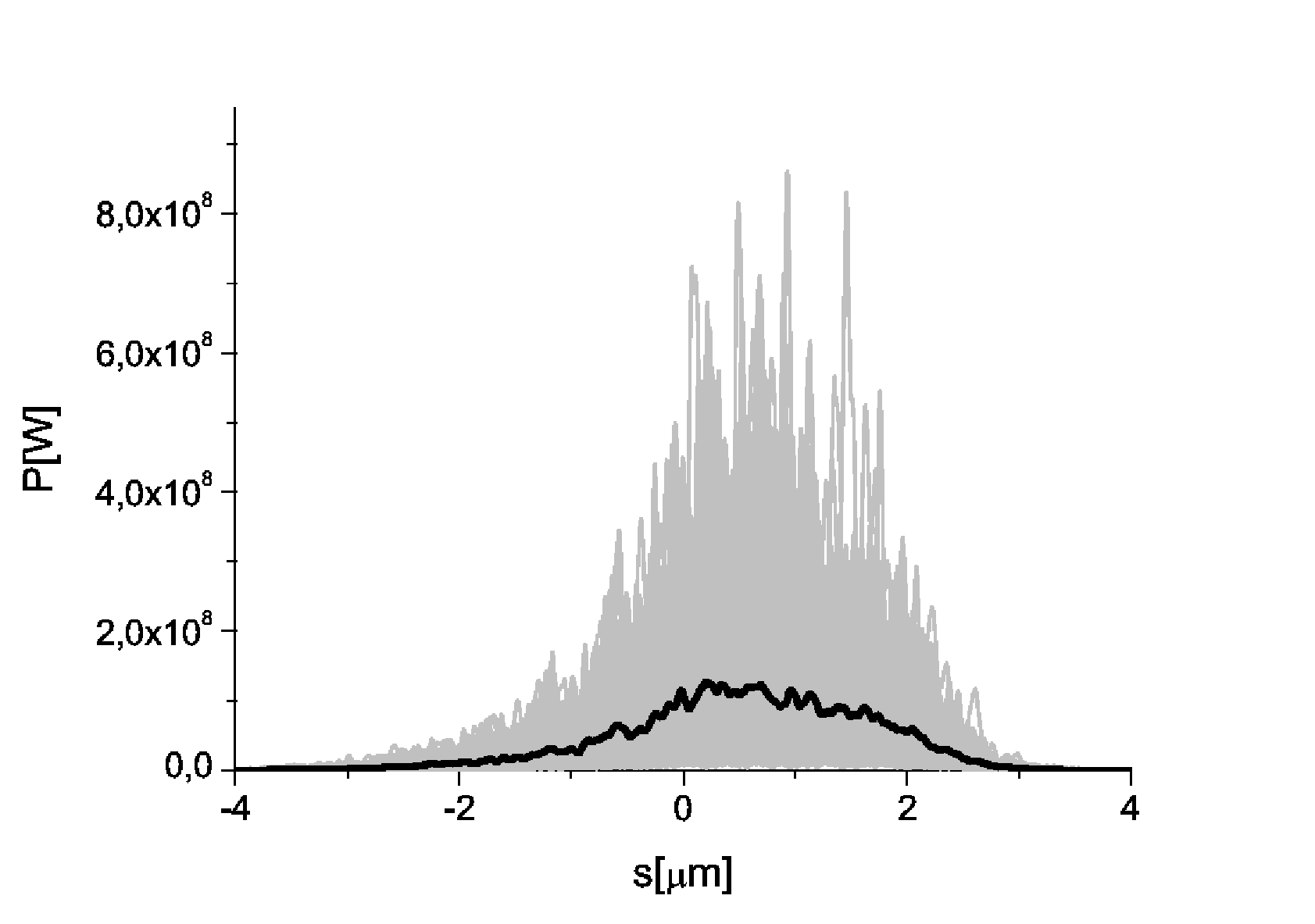}
\includegraphics[width=0.50\textwidth]{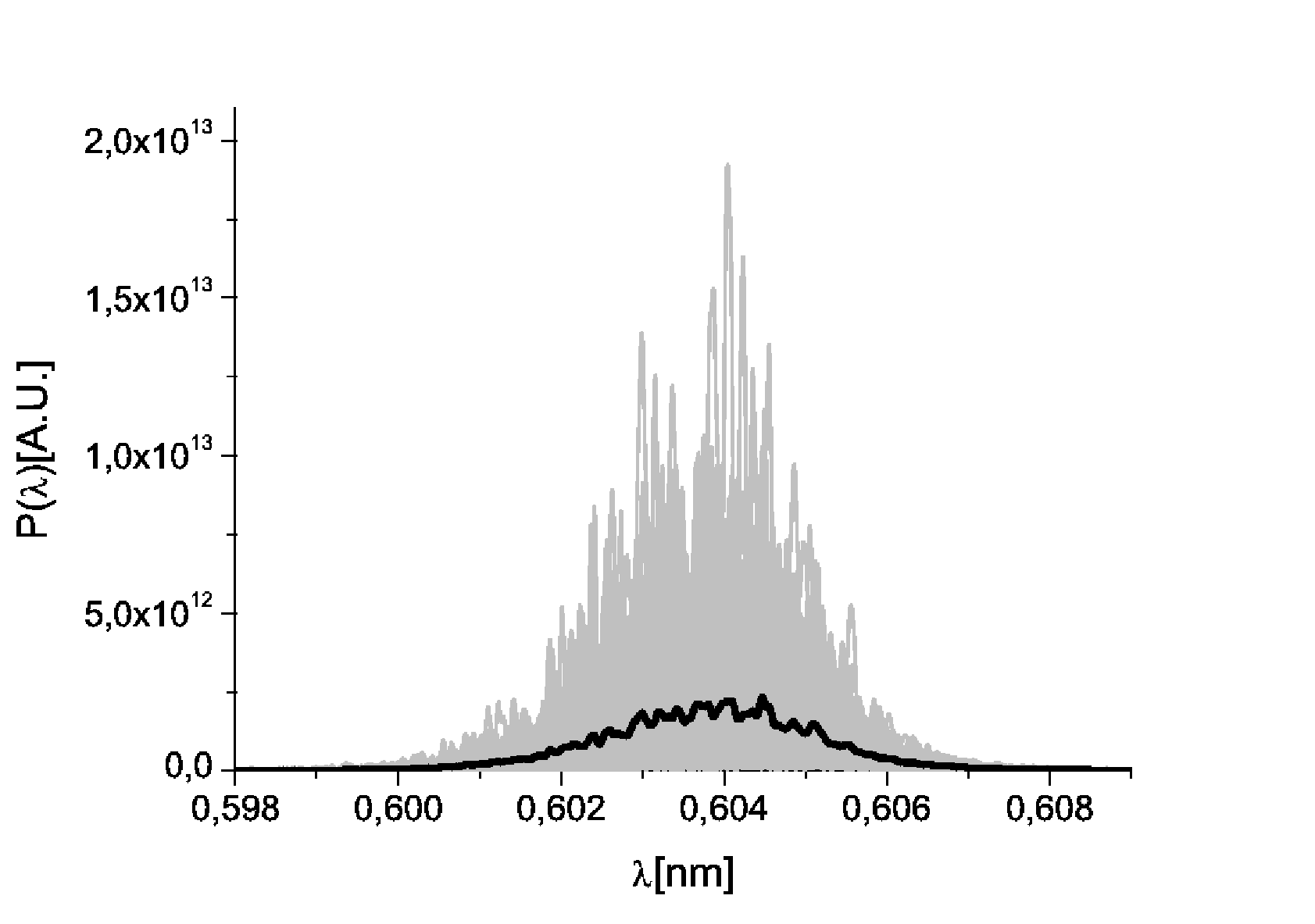}
\caption{Power distribution and spectrum of the SASE soft x-ray
radiation pulse at the exit of the first undulator part U1. Grey
lines refer to single shot realizations, the black line refers to
the average over a hundred realizations.} \label{PSPinS}
\end{figure}
\begin{figure}
\includegraphics[width=0.50\textwidth]{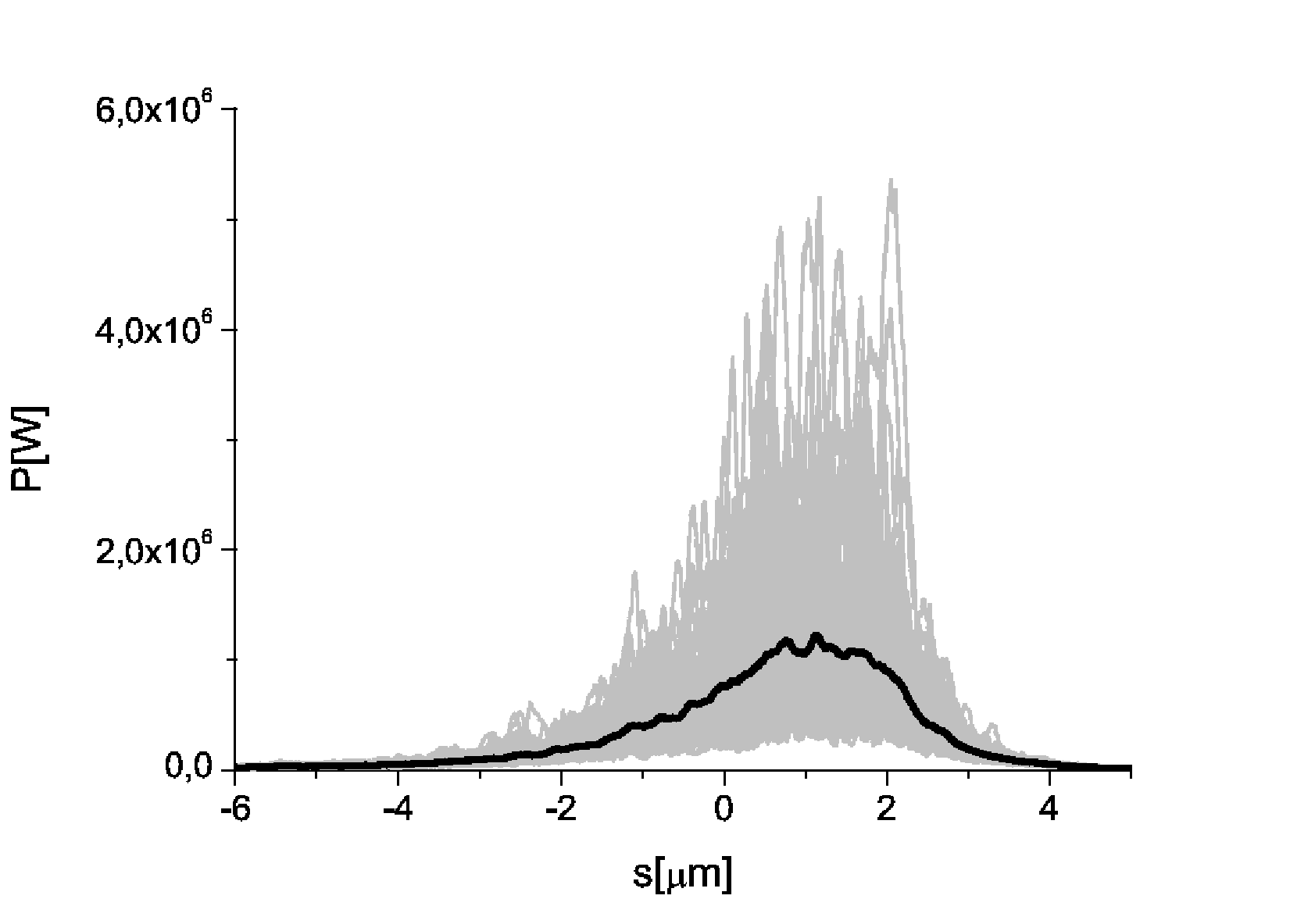}
\includegraphics[width=0.50\textwidth]{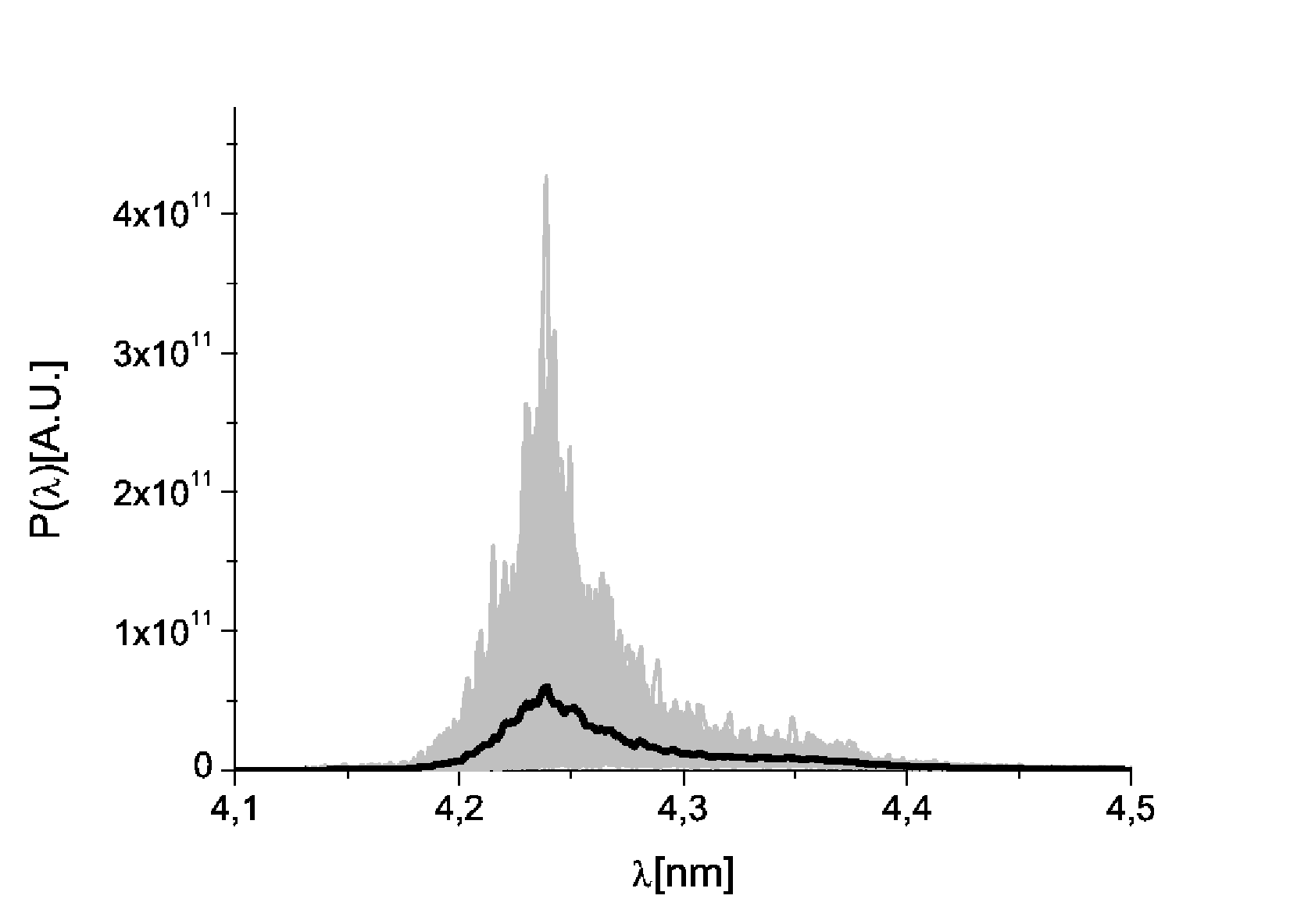}
\includegraphics[width=0.50\textwidth]{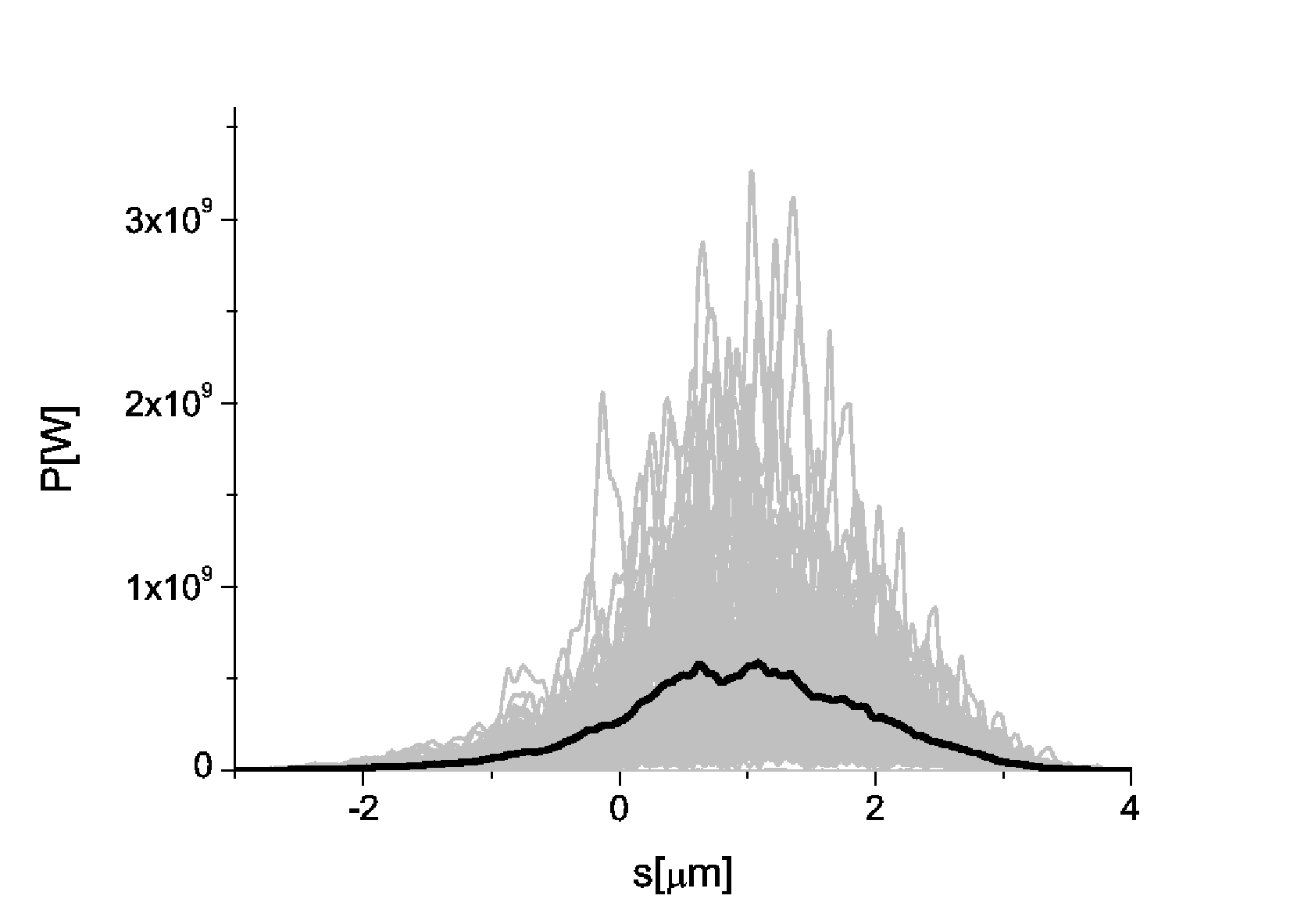}
\includegraphics[width=0.50\textwidth]{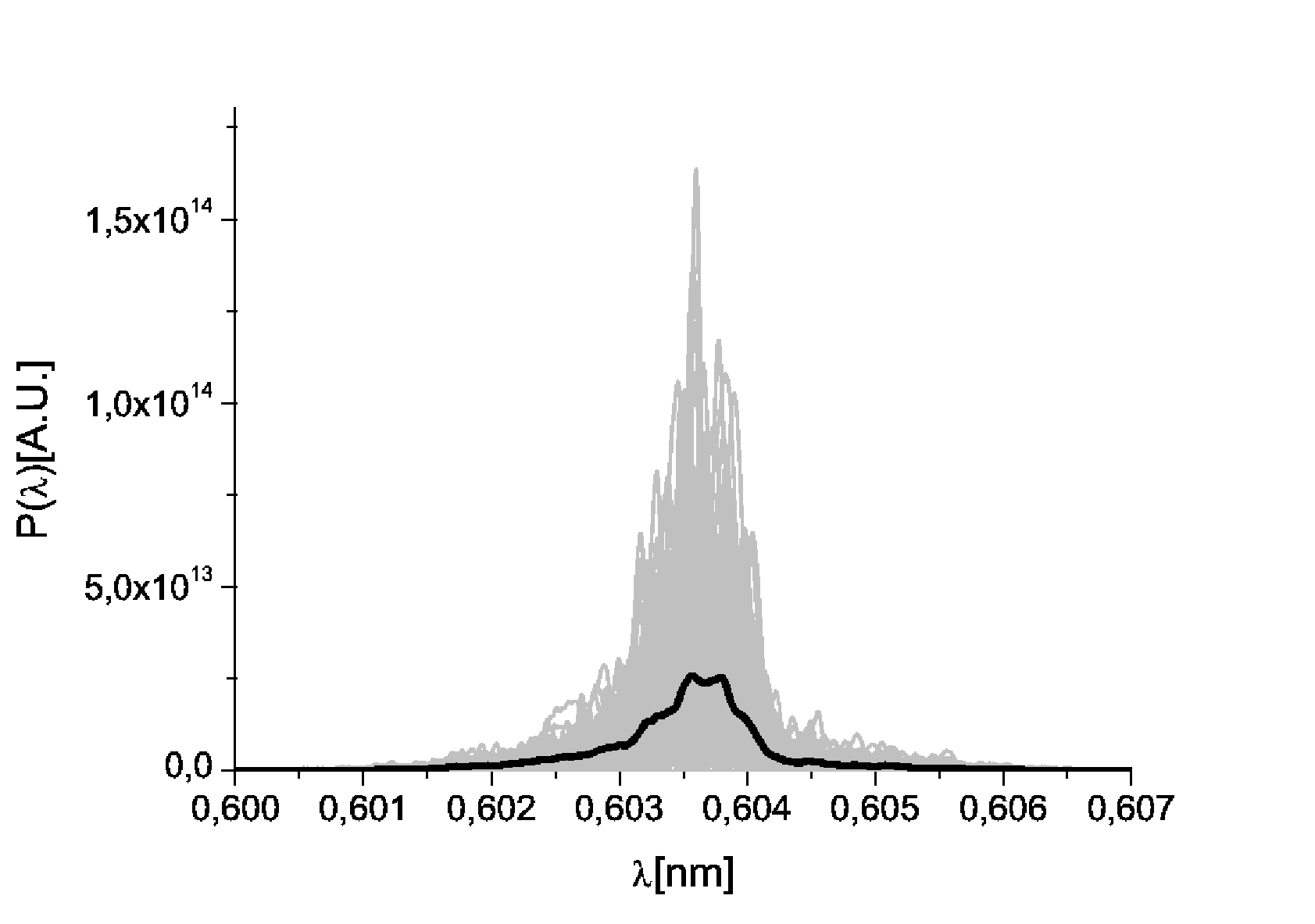}
\caption{SASE radiation power and spectrum at the exit of the second
undulator part U2 (slippage-boosted section). The SASE radiation
generated in U1 is purified in U2, which consists of $2$ cells
resonant at $4.2$ nm. The fundamental radiation at $4.2$ nm is
seeded by shot noise. The harmonic radiation is seeded by that
produced in U1. Top row: Results of numerical simulations for
radiation at the fundamental produced in U2. Bottom row: Results of
numerical simulations for harmonic radiation amplified in U2. Grey
lines refer to single shot realizations, the black line refers to
the average over a hundred realizations.} \label{PSPU2S}
\end{figure}
As it was previously remarked, the number of cells in the undulator
U1 should be equal to five in order to optimize the final
characteristics of the radiation pulse. The output power and
spectrum after the first undulator tuned to $0.6$ nm (the
corresponding rms K value is $2.54$) is shown in Fig. \ref{PSPinS}
for 100 runs. The average behavior is rendered in black. The
radiation field is first dumped at the exit of U1, and then further
imported in the Genesis code for simulating the 7th harmonic
interaction in U2, which is resonant at a fundamental of $4.2$ nm.
Together with the radiation pulse, also electron beam file generated
using the values of energy loss and energy spread at the exit of U1
is fed in the simulation of the second undulator part. The Genesis
7th harmonic field and particle file were downloaded at the exit of
the U2 undulator and used as input file for the Genesis simulations
of the U3 undulator. As explained in the previous section, the
length of the booster U2 is chosen to make sure that the FEL power
at the fundamental wavelength is much lower than that at the chosen
harmonic. The output power and spectrum of fundamental and harmonic
radiation pulse after the U2 undulator tuned to $4.2$ nm (the
corresponding rms K value is $7.16$, and can be achieved by reducing
the undulator gap), that is the seventh subharmonic, are shown in
the left and right plot of Fig. \ref{PSPU2S}. Since the FEL power at
the fundamental wavelength of $4.2$ nm, which is about $1$ MW, is
much lower than that at $0.6$ nm, which is about $1$ GW), phase
shifters are not needed to suppress the lasing at fundamental
harmonic.

\begin{figure}
\includegraphics[width=0.50\textwidth]{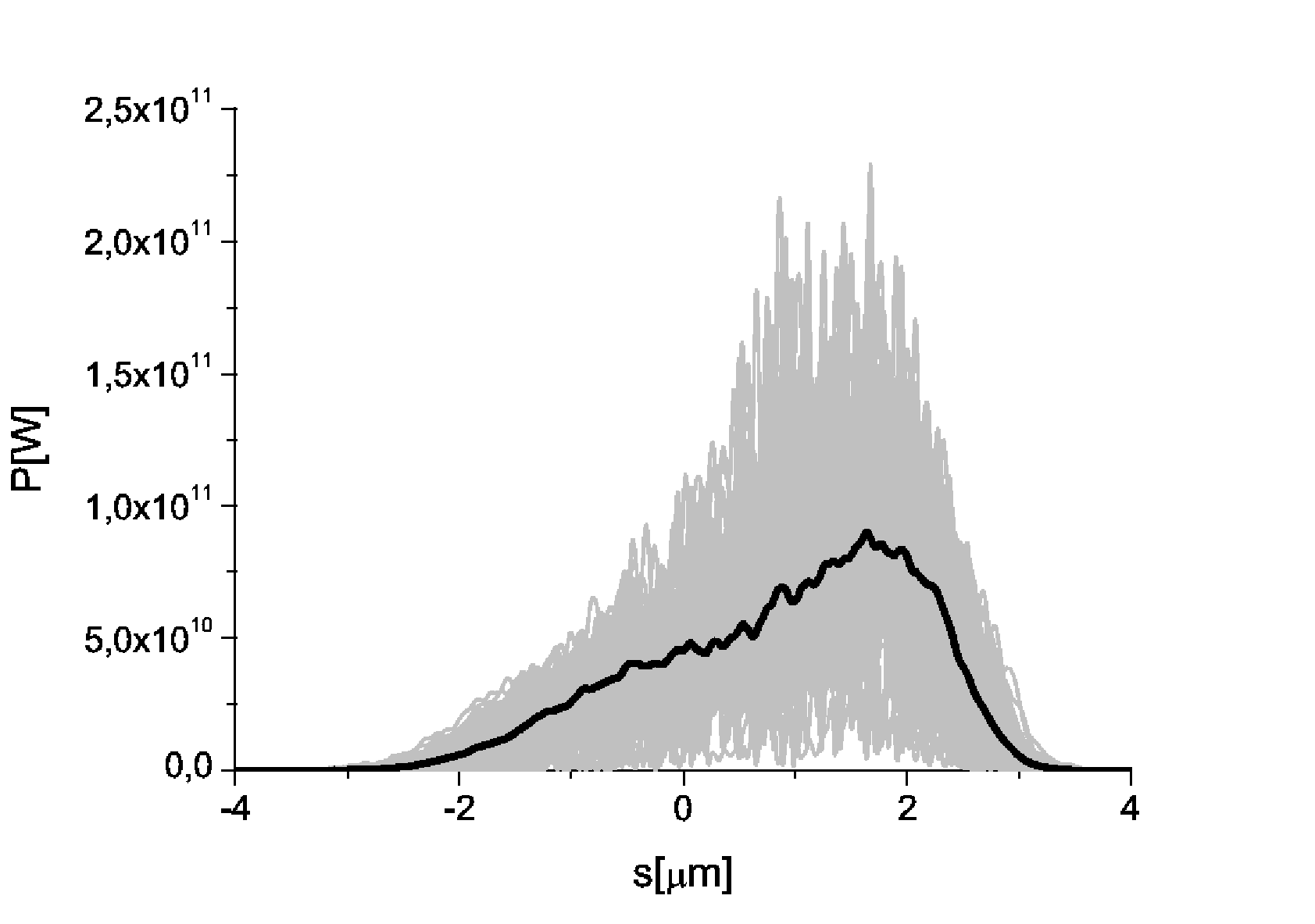}
\includegraphics[width=0.50\textwidth]{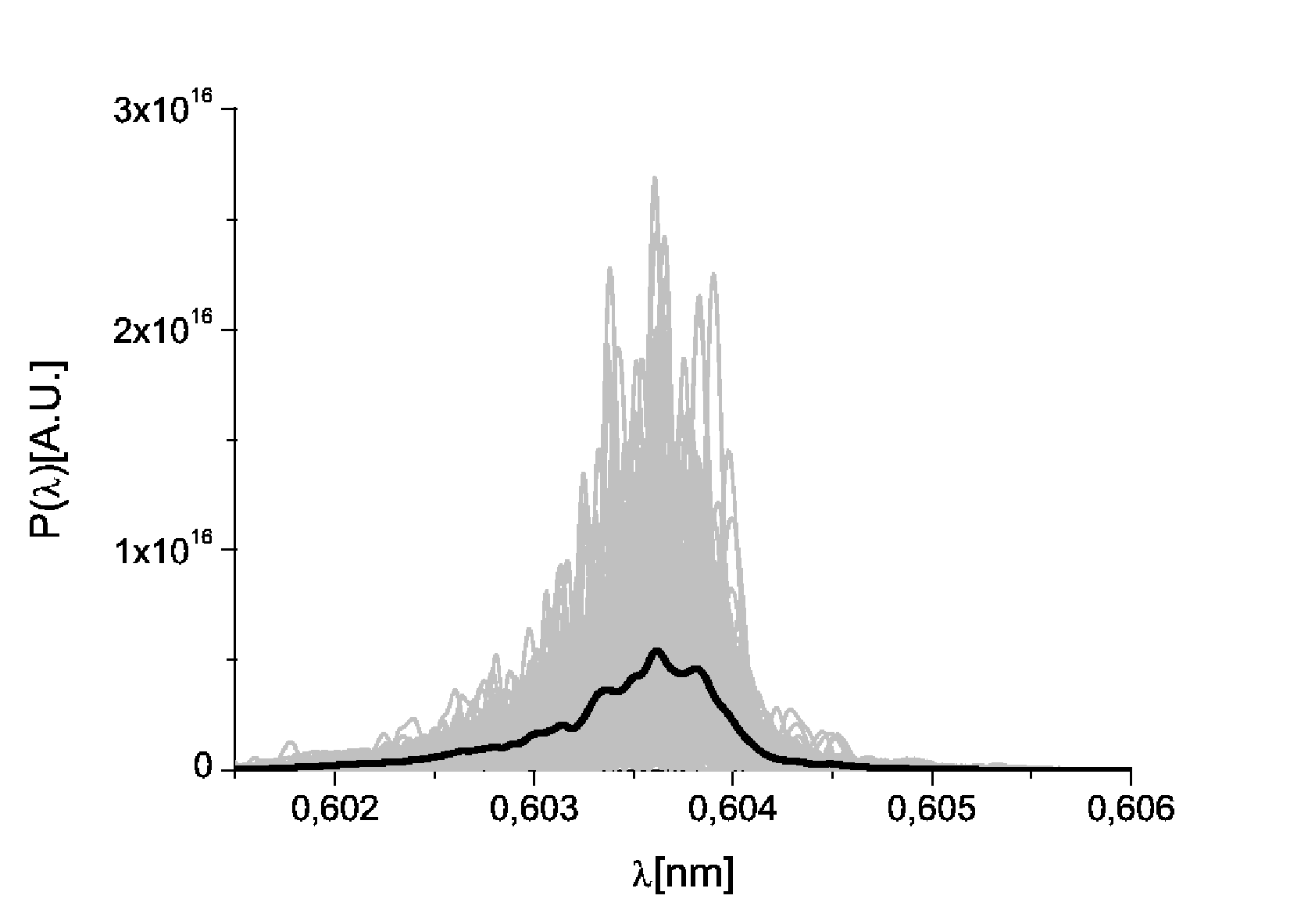}
\includegraphics[width=0.50\textwidth]{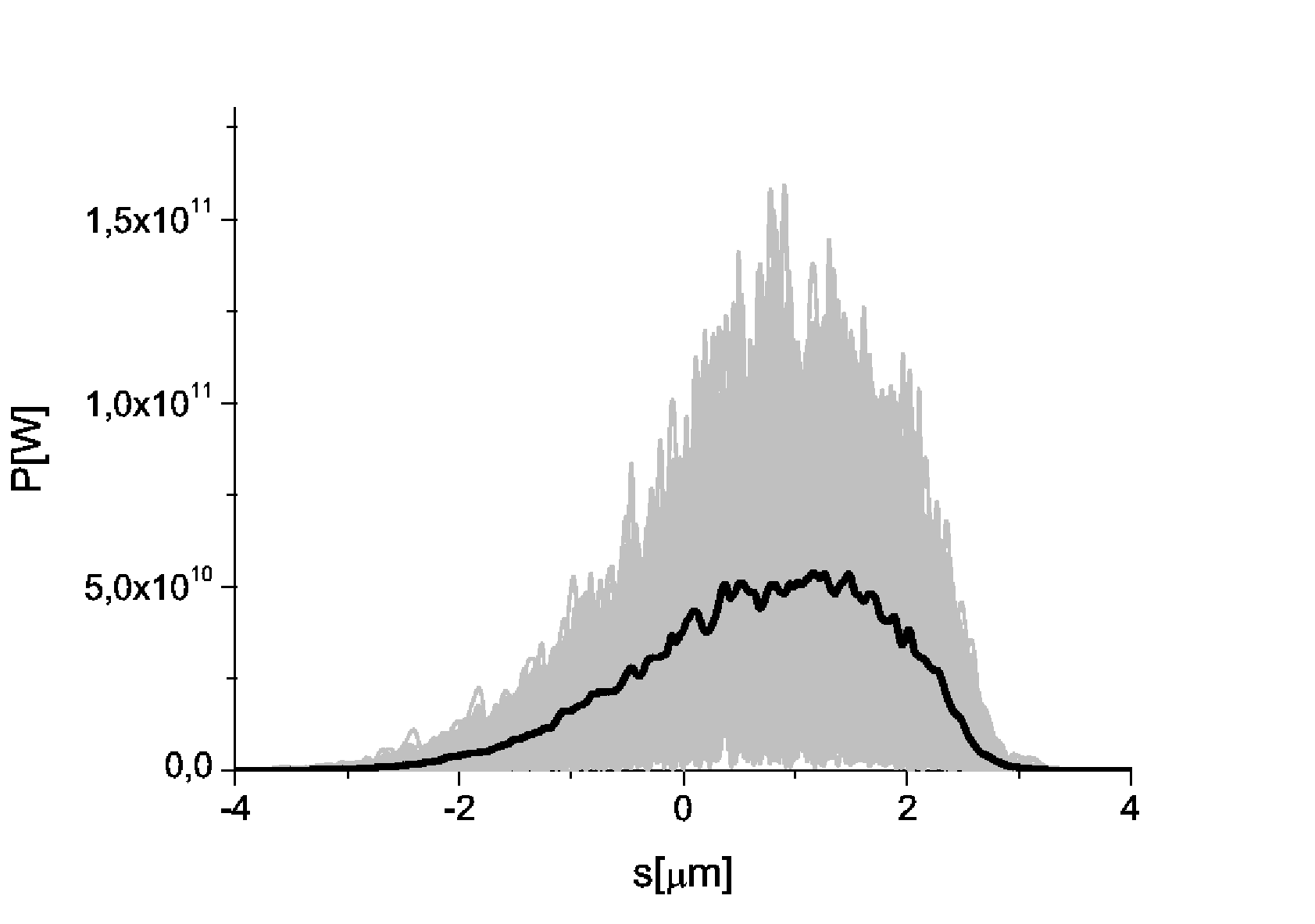}
\includegraphics[width=0.50\textwidth]{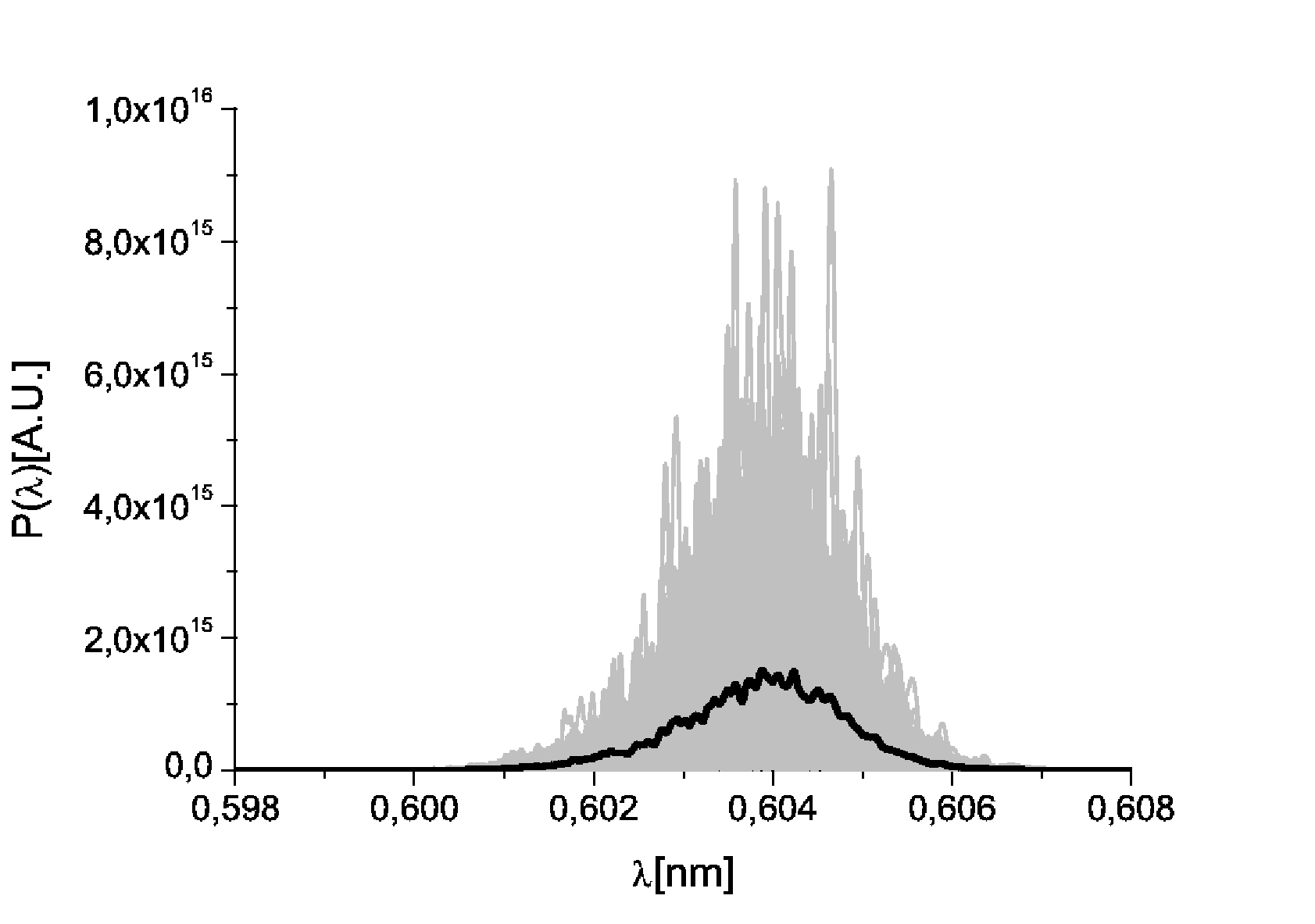}
\caption{Power and spectrum produced in the pSASE mode (top row) and
in the standard SASE mode (bottom row) at saturation without
undulator tapering. Scales of spectral energy density are the same
for both cases. Grey lines refer to single shot realizations, the
black line refers to the average over a hundred realizations.}
\label{PSPoutsatS}
\end{figure}

\begin{figure}[tb]
\includegraphics[width=0.5\textwidth]{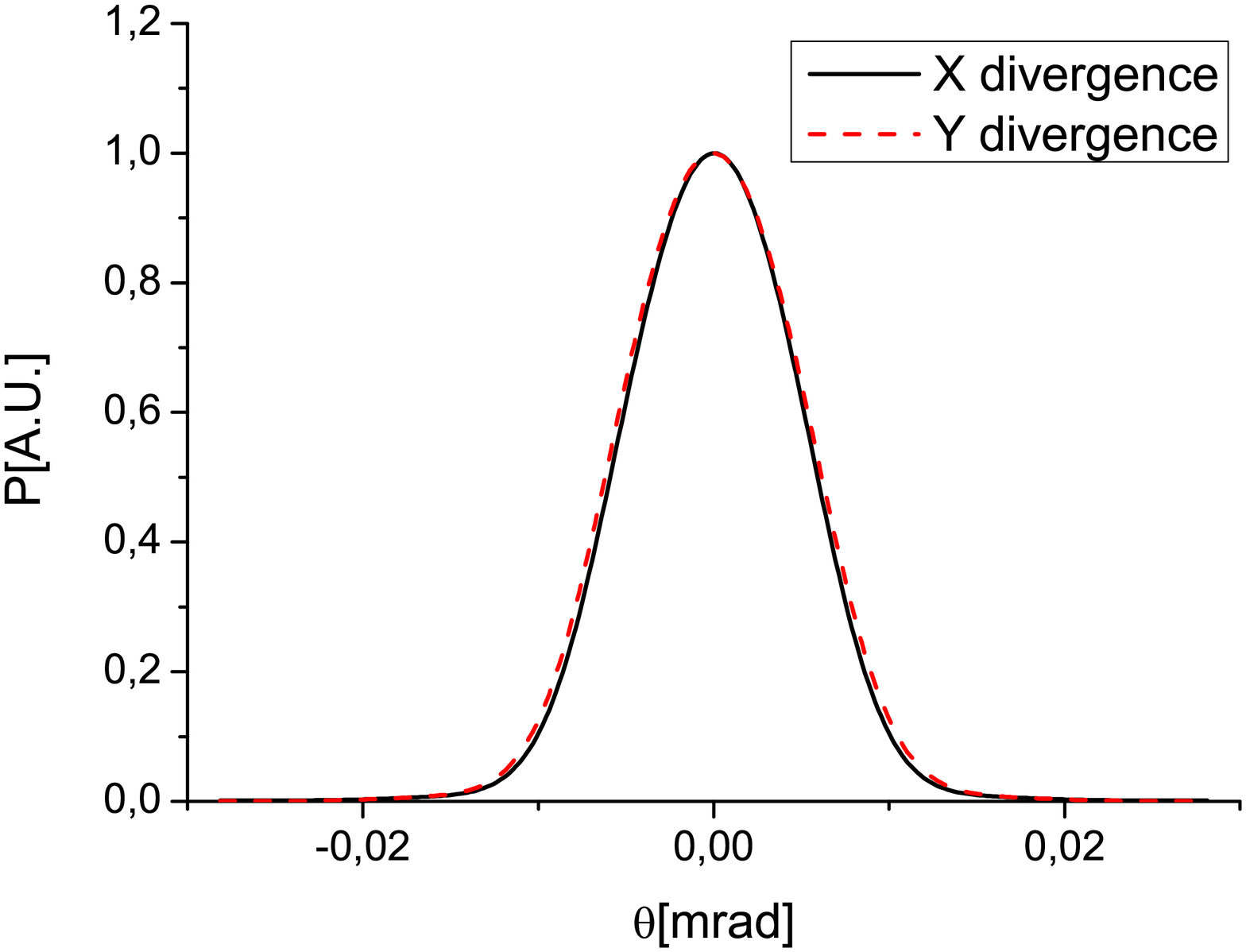}
\includegraphics[width=0.5\textwidth]{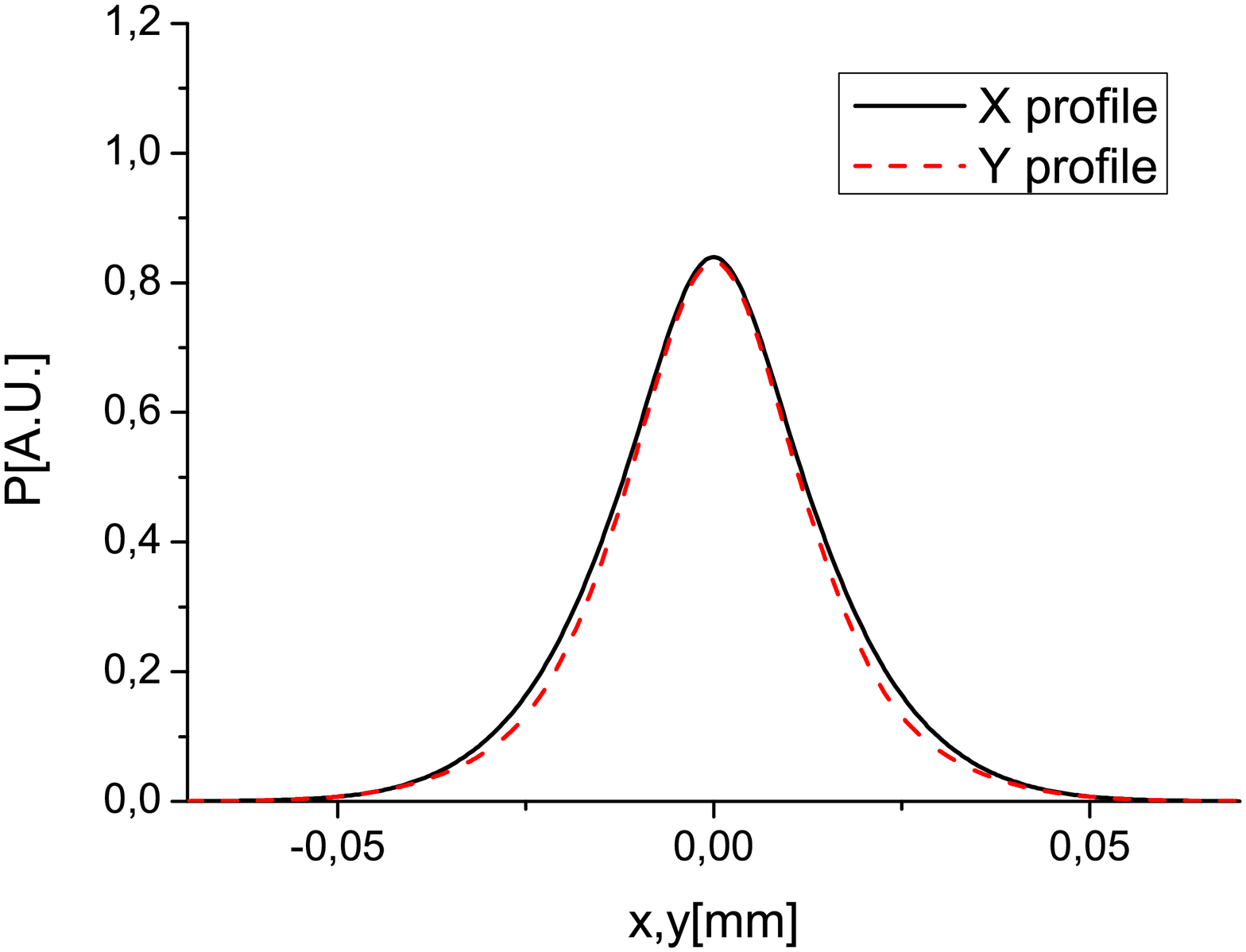}
\includegraphics[width=0.5\textwidth]{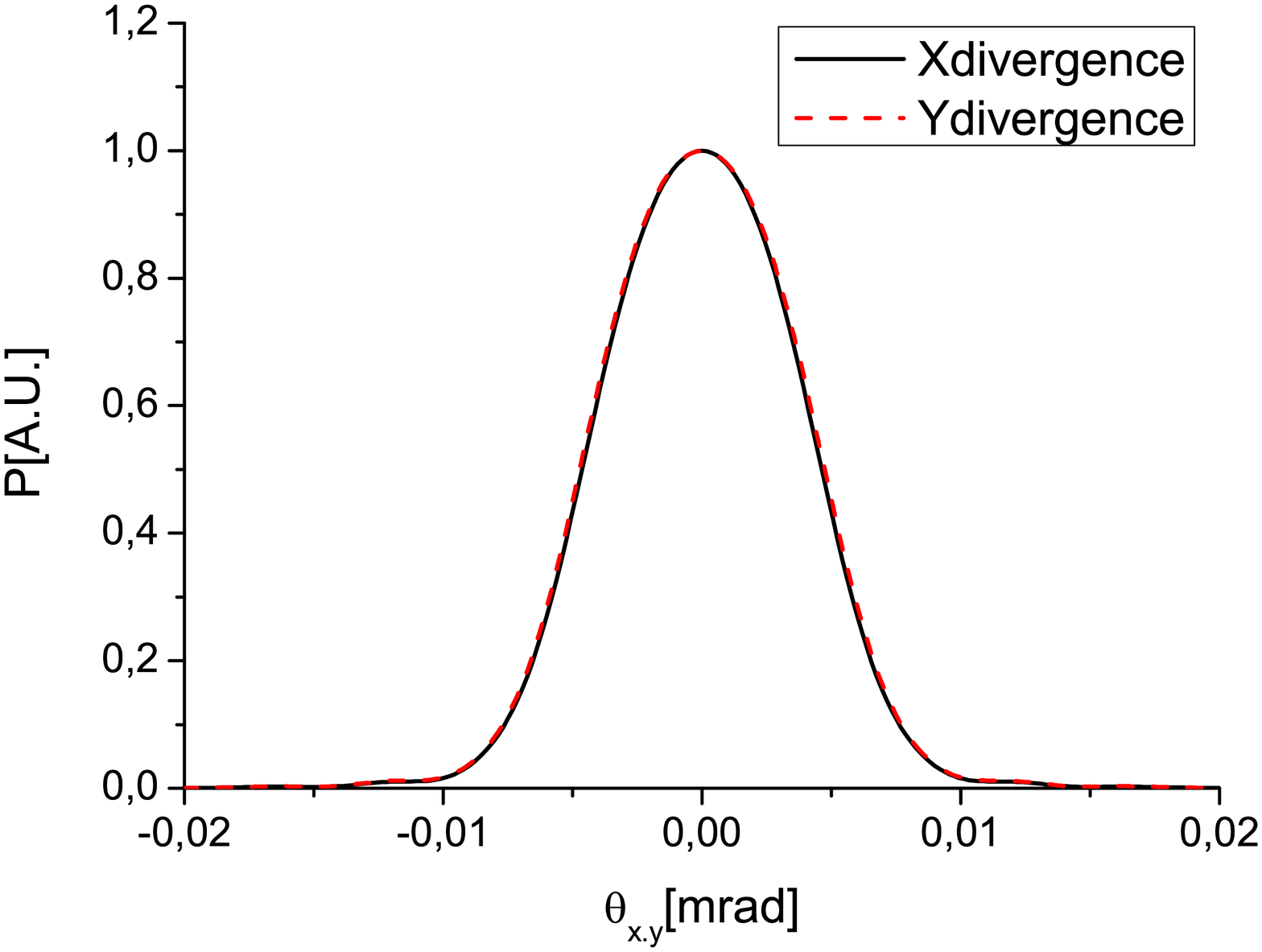}
\includegraphics[width=0.5\textwidth]{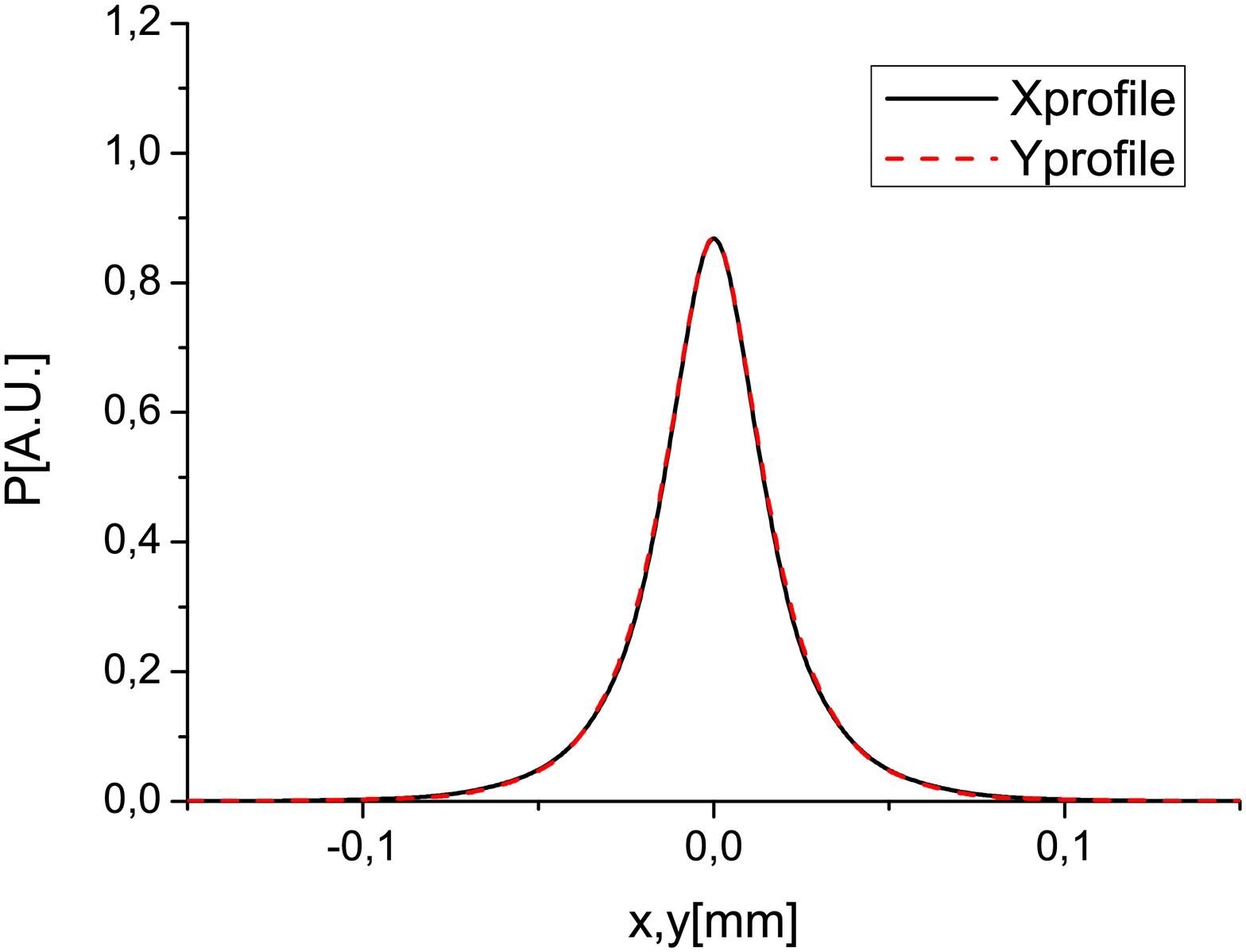}
\caption{Distribution of the radiation pulse energy per unit surface
and angular distribution of the pSASE radiation pulse energy at
saturation (top row) and at the exit of the setup, including
tapering (bottom row).} \label{spotTS}
\end{figure}

\begin{figure}
\begin{center}
\includegraphics[width=0.50\textwidth]{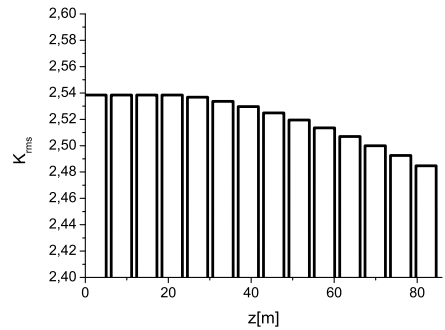}
\end{center}
\caption{Tapering law.} \label{Taplaw}
\end{figure}
\begin{figure}
\includegraphics[width=0.50\textwidth]{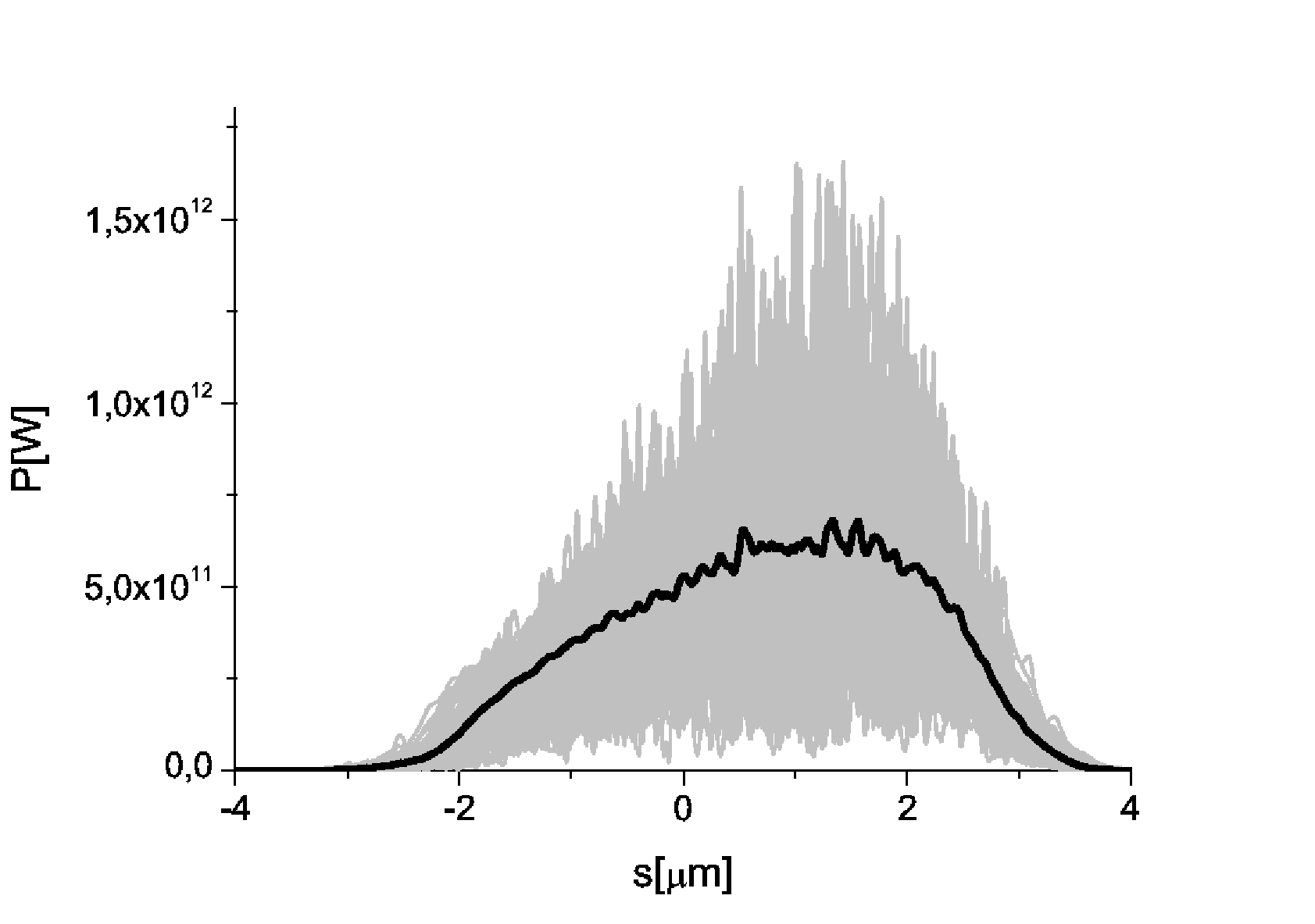}
\includegraphics[width=0.50\textwidth]{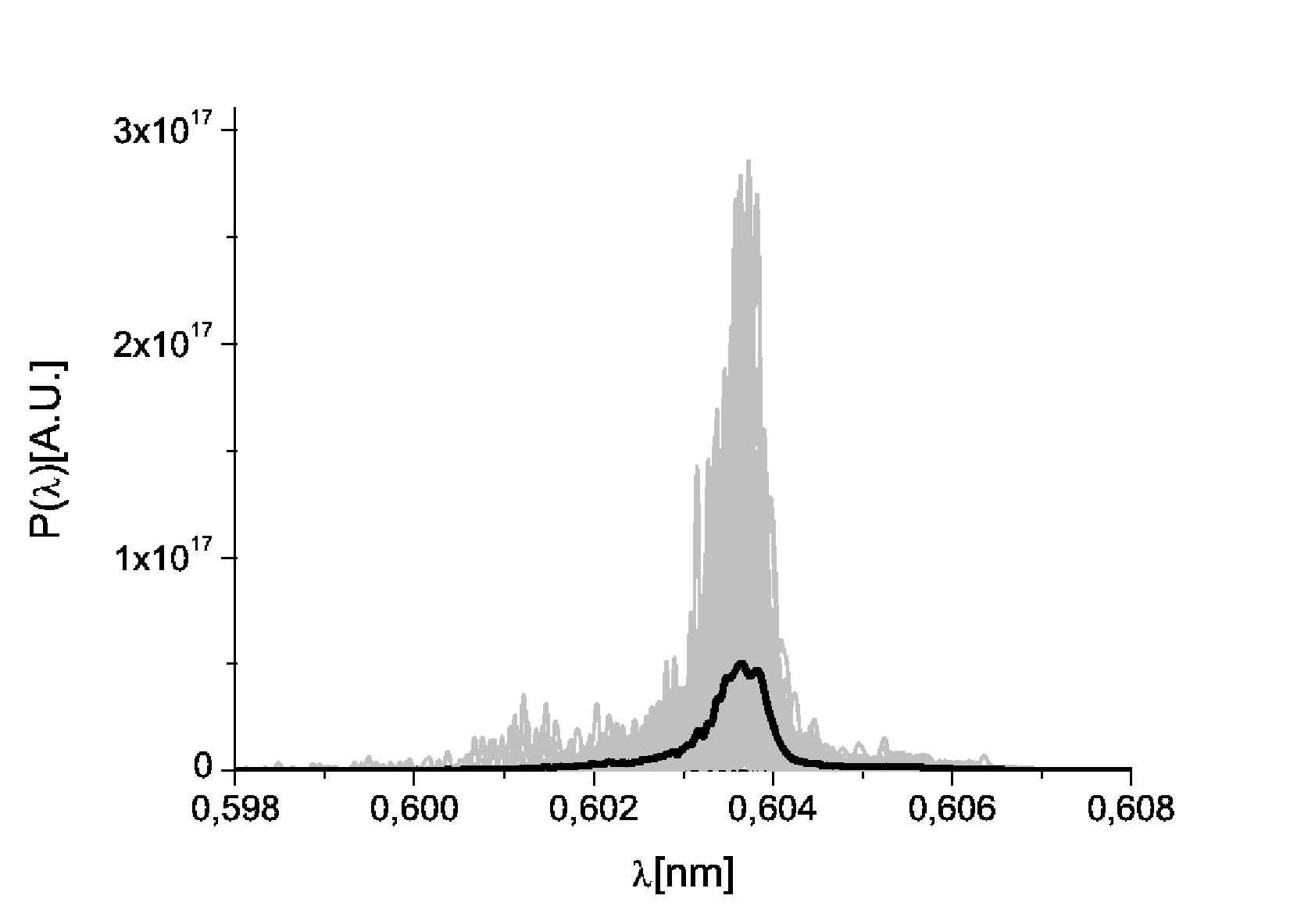}
\caption{Power distribution and spectrum of the purified SASE soft
x-ray radiation pulse at the exit of the setup, with tapering. Grey
lines refer to single shot realizations, the black line refers to
the average over a hundred realizations.} \label{PSPoutapS}
\end{figure}

\begin{figure}
\includegraphics[width=0.50\textwidth]{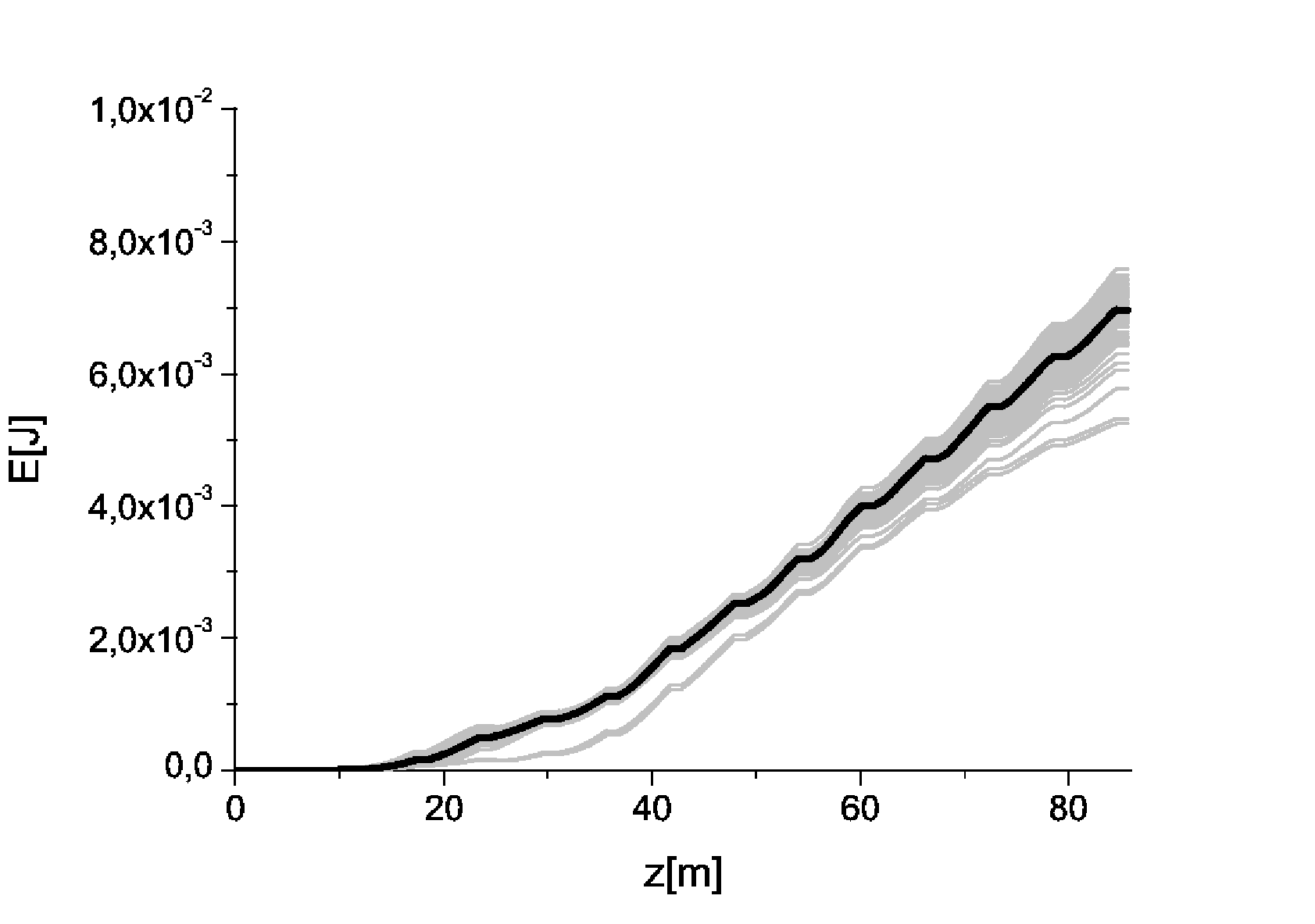}
\includegraphics[width=0.50\textwidth]{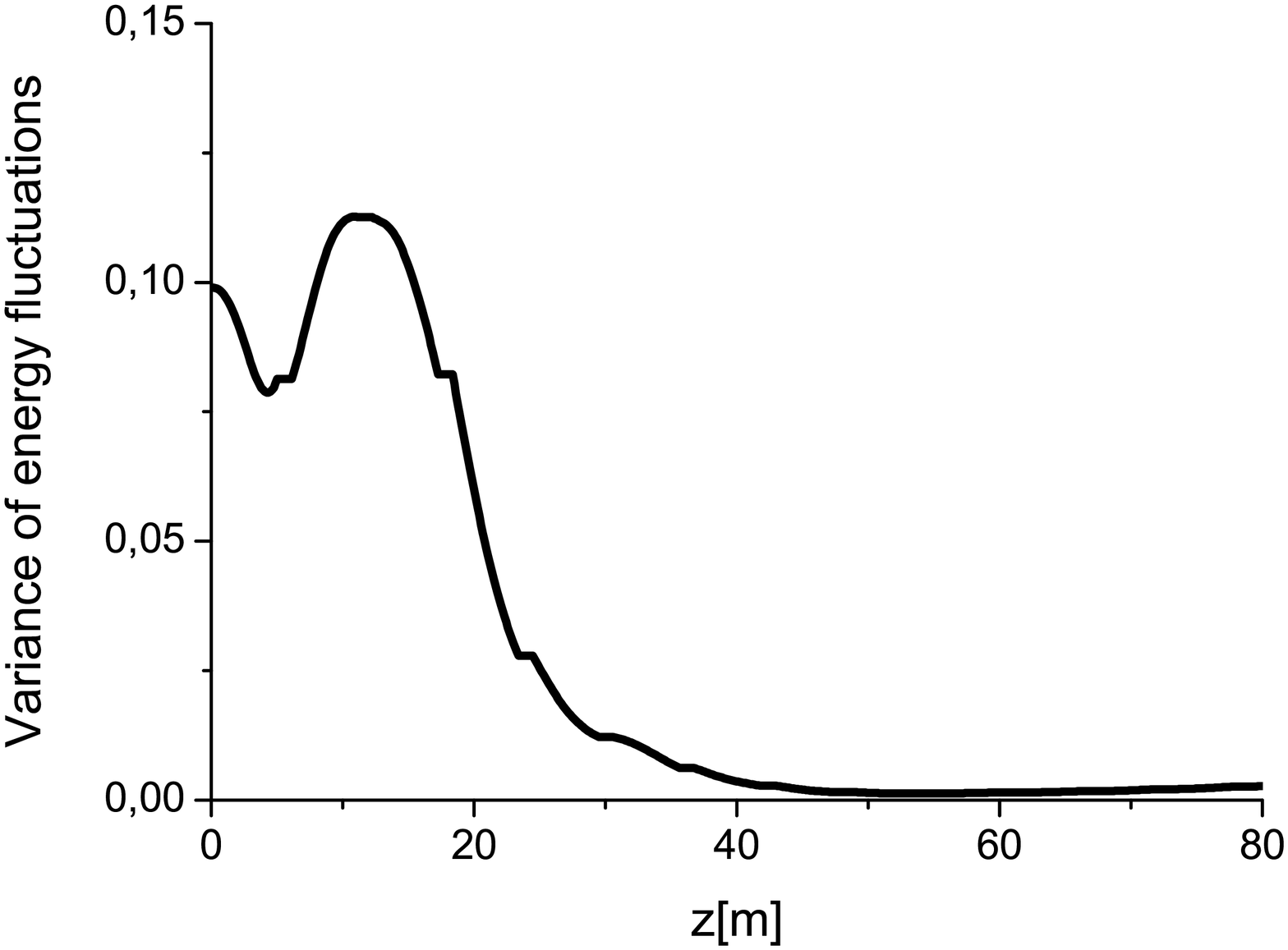}
\caption{Evolution of the output energy in the photon pulse and of
the variance of the energy fluctuation as a function of the distance
inside the output undulator, with tapering. Grey lines refer to
single shot realizations, the black line refers to the average over
a hundred realizations.} \label{EnvaroutS}
\end{figure}
%

As explained in the previous section, the output undulator U3
consists of two sections. The first section is composed by an
uniform undulator, the second section by a tapered undulator. The
purified pulse is exponentially amplified passing through the first
uniform part of the output undulator. This section is long enough,
$5$ cells, in order to reach saturation, which yields about 50 GW
power Fig. \ref{PSPoutsatS} (top row). The radiation power profile
and spectra for SASE3 undulator beamline working in the nominal SASE
mode is shown in Fig. \ref{PSPoutsatS} (bottom row). As seen before,
the power level for both modes of operation are similar, but the
spectral density for the pSASE case is significantly higher than for
the nominal SASE case. The size and divergence of the pSASE
radiation pulse at saturation are shown in Fig. \ref{spotTS} (top
row).

In the second part of the output undulator U3, the purified FEL
output is enhanced up to about $0.6$ TW taking advantage of a taper
of the undulator magnetic field over the last $9$ cells after
saturation. The tapering law is shown in Fig. \ref{Taplaw}. The
output power and spectrum of the entire setup, at the exit of U3, is
shown in Fig. \ref{PSPoutapS}. The size and divergence of the pSASE
radiation pulse at the exit of the setup including undulator
tapering are shown in the bottom row of Fig. \ref{spotTS}. By
inspection, one can see that the difference with the pSASE setup at
saturation, shown in the top row of the same figure, is minimal. The
evolution of the output energy in the photon pulse as a function of
the distance inside the output undulator is reported in Fig.
\ref{EnvaroutS}. As reported in the previous section, the photon
spectral density for the output TW-level pulse is about $30$ times
higher than that for the nominal SASE pulse at saturation.

\section{\label{sec:cons} Conclusions}

We studied the simple scheme proposed in \cite{PSASE} to
significantly enhance the spectral brightness of a SASE FEL with the
help of numerical simulations. Using the parameters for the soft
x-ray beamline SASE3 at the European XFEL we show, using the nominal
electron bunch parameter set, that the SASE bandwidth at saturation
can be reduced by a factor of five with respect to the proposed
configuration of the baseline, variable gap SASE3 undulator, Fig.
\ref{pSASE}. In addition to the example studied in \cite{PSASE}, the
purified radiation after saturation is further significantly
amplified (we report an order of magnitude increase in power) in the
last tapered part of SASE3 undulator. With this configuration, a
pSASE FEL reaches TW peak power level with significantly enhanced
brightness (about two orders of magnitude) compared with the nominal
SASE regime \cite{TSCH}.

\section{Acknowledgements}

We are grateful to Massimo Altarelli, Reinhard Brinkmann, Henry
Chapman, Janos Hajdu, Viktor Lamzin, Serguei Molodtsov and Edgar
Weckert for their support and their interest during the compilation
of this work. We acknowledge useful discussions with Haixao Deng,
Yuantao Ding, Zhirong Huang and Dao Xiang, who shared their
experience in simulating the pSASE configuration with us.

\end{document}